\documentclass[usenatbib]{mnras}
\usepackage{mathptmx}
\usepackage{breakurl}
\usepackage{comment}
\usepackage{longtable}
\usepackage[T1]{fontenc}
\usepackage{bold-extra}
\usepackage{multirow}
\usepackage{graphicx}   
\usepackage{amsmath}   
\usepackage{graphicx}	
\usepackage{amssymb}	
\newcommand{\msun} {M$_{\odot}$}
\newcommand{\src} {NGC\,247 ULX-1}
\newcommand{\xmm} {\textit{XMM-Newton}}

\newcommand{\degmark}{$^\circ$}
\newcommand{\rsph}{$R_{\textrm{sph}}$}
\def \ergsec{\hbox{erg s$^{-1}$}}

\def \arcmin {\hbox{$^\prime$}}
\def \arcsec {\hbox{$^{\prime\prime}$}}

\defcitealias{pinto21}{Paper~I}
\defcitealias{alston21}{Paper~II}

\title[State transitions and dips in NGC 247 ULX-1]
{The chameleon on the branches: spectral state transition and dips in NGC 247 ULX-1}
\author[A. D'A\`i et al.]{A.~D'A\`i$^{1}$, \thanks{E-mail: antonino.dai@inaf.it} 
C. Pinto$^{1}$,
M. Del Santo$^{1}$, 
F. Pintore$^{1}$, 
R. Soria$^{2,3}$,
A. Robba$^{1,4}$, \and
E. Ambrosi$^{1}$, 
W. Alston$^{5}$, 
D. Barret$^{6}$,
A.C. Fabian$^{7}$, 
F. F\"urst$^{8}$,
E. Kara$^{9}$, \and
P. Kosec$^{9}$,
M. Middleton$^{10}$,
T. Roberts$^{11}$,
G. Rodriguez-Castillo$^{1}$,
D. J. Walton$^{7}$
\vspace{6pt}\\
$^{1}$ INAF/IASF Palermo, via Ugo La Malfa 153, I-90146 - Palermo, Italy\\
$^{2}$ College of Astronomy and Space Sciences, University of the Chinese Academy of Sciences, Beijing 100049, China\\
$^{3}$ Sydney Institute for Astronomy, School of Physics A28, The University of Sydney, Sydney, NSW 2006, Australia\\
$^{4}$ Università degli studi di Palermo, Dipartimento di Fisica e Chimica, via Archirafi 36, I-90123 Palermo, Italy\\
$^{5}$ European Space Agency (ESA), European Space Astronomy Centre (ESAC), E-28691 Villanueva de la Ca\~{n}ada, Madrid, Spain\\
$^{6}$ CNRS, IRAP, Université de Toulouse, 9 Avenue du colonel Roche, BP 44346, F-31028, Toulouse, Cedex 4, France\\
$^{7}$ Institute of Astronomy, Madingley Road, CB3 0HA Cambridge, United Kingdom\\
$^{8}$ Quasar Science Resources for ESA, European Space Astronomy Centre (ESAC), E-28691 Villanueva de la Ca\~{n}ada, Madrid, Spain\\
$^{9}$ MIT Kavli Institute for Astrophysics and Space Research, Cambridge, MA 02139, USA\\
$^{10}$ School of Physics \& Astronomy, University of Southampton, Southampton, SO17 1BJ, UK\\
$^{11}$ Centre for Extragalactic Astronomy \& Dept of Physics, Durham University, South Road, Durham DH1 3LE, UK\\
}

\date{Accepted XXX. Received YYY; in original form ZZZ}
\pubyear{2021}

\begin{document}
\label{firstpage}
\pagerange{\pageref{firstpage}--\pageref{lastpage}}
\maketitle

\begin{abstract}
Soft Ultra-Luminous X-ray (ULXs) sources are a sub-class of the ULXs that can switch from a supersoft spectral state, where most of the luminosity is emitted below 1 keV, to a soft spectral state with signiﬁcant emission above 1 keV. In a few systems, dips have been observed. The mechanism behind this state transition and the dips nature are still debated. To investigate these issues, we obtained a long \xmm\ monitoring campaign of a member of this class, \src. We computed the hardness-intensity diagram for the whole data-set and identiﬁed two different branches: the normal branch and the dipping branch, which we study with four and three hardness-intensity resolved spectra, respectively. All seven spectra are well described by two thermal components: a colder ($kT_{\rm bb}$  $\sim$ 0.1–0.2 keV) black-body, interpreted as emission from the photo-sphere of a radiatively-driven wind, and a hotter ($kT_{\rm disk} \sim 0.6$ keV) multi-colour disk black-body, likely due to reprocessing of radiation emitted from the innermost regions. In addition, a complex pattern of emission and absorption lines has been taken into account based on previous high-resolution spectroscopic results. We studied the evolution of spectral parameters and ﬂux of the two thermal components along the two branches and discuss two scenarios possibly connecting the state transition and the dipping phenomenon. One is based on geometrical occultation of the emitting regions, the other invokes the onset of a propeller effect.

\end{abstract}

\begin{keywords}
X-rays: binaries -- X-rays: individual: \src. 
\noindent
Facility: {\it XMM-Newton}
\end{keywords}

\section{Introduction} \label{sect:intro}
{\textit{Ultra-luminous X-ray sources}} (ULXs; see \citealt{Kaaret2017} and references therein) are  point-like, off-nuclear, X-ray sources with isotropic X-ray luminosity  above $10^{39}$ erg/s, which is the Eddington limit for a 10\,\msun\ black-hole (BH). Nowadays, it is widely accepted that they are mainly stellar-mass compact objects, such as neutron stars (NS) or BHs, accreting matter from a donor above the Eddington limit. Recent discoveries in this field are refining our understanding: coherent pulsations show that NSs are able to accrete at super-Eddington ratios; relativistic winds provided a spectroscopic tool to measure the mass-outflow rate and its effects onto the accretion process; comparative studies on the shape of the broadband continuum in different sources hint to system inclination as an important ingredient; the presence of massive companion stars explain the evolutionary path of these systems towards the well-known classical high-mass systems (see, e.g., \citealt{Bachetti2014, Pinto2016, Walton2018a, Kosec2018b}). 

The X-ray spectra of ULXs can be approximated as power-law with indices in a wide range from ~1 to ~3 in the soft (1-5 keV) band and a cut-off at energies above 10 keV \citep{Bachetti2014, Pinto2016, Walton2018a}. Of particular interest are the so called soft ultraluminous sources (SUL) with spectral indices steeper than $\Gamma$\,=\,2 where the bulk of observed emission is likely associated with re-processing of harder X-rays from the central engine by radiatively-driven optically-thick outflows expected to be present for sources accreting at luminosities exceeding the Eddington limit \citep{Middleton2015a}. This gives an opportunity to test extreme accretion regimes rarely observed in other accreting objects such as Galactic X-ray binaries and active Galactic nuclei.

\textit{Supersoft ultraluminous X-ray sources} (SSUL; also known as ULS in the literature) can be considered a subclass of SUL, showing bolometric luminosities $L_{\rm bol}$ close to, or in excess of 10$^{40}$ erg s$^{-1}$, but their spectra are completely dominated by black-body emission at a temperature of $\approx$50--140 eV  and associated black-body radii of $\approx$5--10 $\times10^{4}$ km, with little emission above 1 keV \citep{Urquhart2016a}. They were initially interpreted as a completely different physical class of sources. 
It was even suggested that the thermal spectrum could be represented by a multi-colour emission from a standard Shakura-Sunyaev disk \citep{shakura73} in the sub-Eddington high/soft state of an accreting BH. 
In that scenario, the low peak temperature and large radius would be consistent with long-sought-for intermediate mass BHs
(e.g. for \src\ \citealt{tao12} estimated $M_{\rm BH} \sim 600\,$\msun).

However, from the spectral analysis of a homogeneous sample of SSUL sources, \citet{soria16} and \citet{Urquhart2016a} showed that the temperature-luminosity dependence did not follow the typical $L \propto T^4$ law, expected for sub-Eddington accretion onto a standard Shakura-Sunayev disk, nor the $L \propto T^2$ trend expected for a near-Eddington slim-disk. Instead, they showed an approximate $R \propto T^{-2}$ trend, both for individual sources and across the sample, which argues in favour of emission from an extended, optically-thick, photosphere just outside the spherization radius ($R_{\textrm{sph}}$), where radiation pressure launches the densest outflows \citep{poutanen07}. In this scenario, SSULs are considered as ULXs in which the hard photons have been completely reprocessed and thermalised (or, at least, are completely masked from our view). The radial location (size) of the thermal photosphere depends on the physics of the thick inflow/outflow, as well as on the polar angle \citep{Takeuchi2013, Middleton2015a}.

To test this SSUL/ULX unification scenario, we planned to identify and study in more detail a source that spends some time in the SSUL state and some time at the soft end of the standard ULX classification. This would give us a chance to monitor, for example, how the transition is associated with changes in the black-body temperature, luminosity, time-variability properties. Such a source exists, in the galaxy NGC\,247, as we describe below.

\subsection{NGC 247 ULX-1}
NGC\,247 is a barred, late-type, spiral galaxy 
at a distance of 3.38\,$\pm$\,0.06 Mpc \citep[estimated with the Cepheids method,][]{gieren09}.
It is viewed almost face-on  and contains a significant population of X-ray point sources, most likely accreting XRBs, whose brightest member is \src\ \citep{jin11}. Although the overall star-forming rate in NGC\,247 is low \citep[$\approx$0.1\,M$_{\odot}$ yr$^{-1}$:][]{davidge06}, \src\ is located (or at least projected) close to an OB stellar association with an estimated age of $\approx$5--40 Myr \citep{tao12}. The spectrum of the bright optical counterpart of \src\ is compatible with a yellow/blue supergiant with a temperature of (19,000\,$\pm$\,400) K and radius of about 60\,$R_{\odot}$ \citep{feng16}. However, a strong caveat is that for most ULX counterparts, it is difficult or impossible to distinguish between the optical emission from the companion star and that from the irradiated accretion disk \citep[][]{grise12, ambrosi18}.

\citet[][]{jin11} reported the first detailed X-ray spectral study of \src\ from an \xmm\ observation performed in 2009. The spectrum could be satisfactorily modelled by a combination of a soft disk blackbody (or a simple blackbody) of  temperature 0.11-0.12 keV, accounting for about 90\% of the total emission, and a power-law component, whose photon index could not be tightly constrained, that accounted for the harder part of X-ray spectrum. A possible absorption feature around 1 keV was also detected. In a new \xmm\ observation in 2014 \citet{feng16} found the source in a different spectral state, where the flux of the power-law component became comparable to the softer component and \src\ appeared for the first time as a standard ULX in the SUL regime. 
Superposed on this continuum, spectral residuals are sometimes observed around 1 keV; they have been interpreted either as absorption lines from an optically thick, photo-ionised wind \citep{jin11, middleton14}, or/and as emission lines from a hot (kT $\sim$1 keV) collisionally-ionised plasma \citep[][]{feng16}. The lack of sufficient energy resolution and the low statistics of those observations have prevented so far an unambiguous determination of the nature of those features, although their presence is generally considered a characteristic property of the SSUL class \citep[see][]{Urquhart2016a}. 

In the 2009 observation, \citet{feng16} found also X-ray dips in \src\ light curve, which seem to be linked to the SUL state occurrence. No increase in the neutral absorption column as a dip occurred was noted, while the continuum appeared generally to soften. 

\subsection{The recent results from the campaign}
With the goal to investigate the nature of the variability, from minutes to hours, in super-soft ULXs we proposed and were awarded a deep \xmm\ programme onto \src\ (PI C.\,Pinto). The main objective is to know whether such variations in ULXs, and particularly in SSUL, are due to stochastic variability in the wind (at a given mass accretion rate $\dot{M}$) or the clumpy nature of the wind, or a change in the underlying $\dot{M}$.

Observations were carried out between 2019-12-03 and 2020-01-12, collecting approximately 800 ks. Given the rich amount of new information derivable from this campaign, the study of these data has been organised into a series of works that focus on different aspects of \src. Two other papers have recently appeared. The first one, \citet[][hereafter Paper I]{pinto21}, searched for evidence of winds mainly thanks to the high-resolution Reflection Grating Spectrometer (RGS) spectra. Two main highly-ionised wind components (one in emission and one in absorption) have been discovered. By applying self-consistent plasma models, the absorbing wind component showed a (model-independent) blue-shift of $\sim 0.17c$, whereas the emission line pattern has a velocity along our line-of-sight of +0.042$c$ or -0.022$c$ depending on the use of collisionally- or photo-ionised plasma models. The second paper, \citet[][hereafter Paper II]{alston21}, focused on the temporal variability and on the occurrence and origin of the dipping behaviour. Some preferential time-scales corresponding to quasi-periodic oscillations (QPOs) at frequencies of 4, 9 and 20 $\times 10^{-5}$ Hz were detected in the power spectrum of observations with dips.

The present paper focuses on the X-ray spectral continuum. In particular, we aim to characterise the broadband spectral variability as a function of the observed accretion state and spectral hardness, constrain the average spectral shape of the source during dip episodes, and understand the occurrence and the spectral changes associated with state transition and the link between the occurrence of dips and the source spectral state. For this work we used data from the European Photon Imaging Camera (EPIC) detectors: the two Metal Oxide Semi-conductor (MOS) cameras and the EPIC-pn, (PN, hereafter) camera. We refer to \citetalias{pinto21} for more details on the whole \xmm\ campaign.

\section{Observations and data reduction}

The NGC 247 campaign comprises eight \xmm\ orbits, with Observation Identifications (ObsID) numbers: 0844860-101/201/301/401/501/601/701/801 (we shall refer to a specific ObsID using these last three digits). All \xmm/EPIC observations were acquired through the medium filter using the full window mode. We used the \textsc{Science Analysis System} (SAS) v. 18.0.0 for the extraction and reduction of the \xmm\ data and \textsc{HeaSOFT} v. 6.28 for data processing and spectral fitting. Spectral parameter uncertainties are estimated at the 68\% confidence level for a single parameter of interest.

We used the tasks \textsc{emproc} and \textsc{epproc} from the  SAS pipeline to process raw data into clean event files. We adopted standard filtering criteria for product generation (FLAG==0 and PATTERN $\leq$ 4 and $\leq$ 12 for PN and MOS, respectively). We extracted source light curves and spectra  using a circular region around the best-known position of \src\ (source coordinates taken from \citealt{tao12}) wih a radius of 32\arcsec.  We selected a similar circular region for the background region from a source close-by area of the same CCD, but away from its read-out direction, approximately at a distance of {1.5\arcmin} from the source coordinates,  where no other point-like source was present. The background regions were chosen for each ObsID, after visual inspection of each EPIC image, to ensure the correct application of the above-mentioned criteria. Similarly, we produced for each ObsID an energy-selected (E $>$ 10 keV) light curve from the source region to remove episodes of high soft protons background according to the standard procedure\footnote{\url{https://www.cosmos.esa.int/web/xmm-newton/sas-thread-epic-filterbackground}}. This gave us $\sim$\,620 ks of clean data (per EPIC camera). We corrected light curves for effective area, dead-time and calibration issues using the \textsc{epiclccorr} task. 

\section{Light curves, hardness-intensity diagram and spectral filtering}
We extracted PN light curves in the 0.2-10 keV energy band using a time bin size of 100\,s and estimated their variability in terms of root mean square (RMS) per observation. Data-sets 301, 701 and 801 show low variability (RMS values < 10\%), whereas ObsIDs 201, 401, 501 and 601 show RMS values > 30\% due to the presence of dips that last from a few hundreds of seconds to several ks. 

\begin{figure*}
\centering
\includegraphics[width=2\columnwidth]{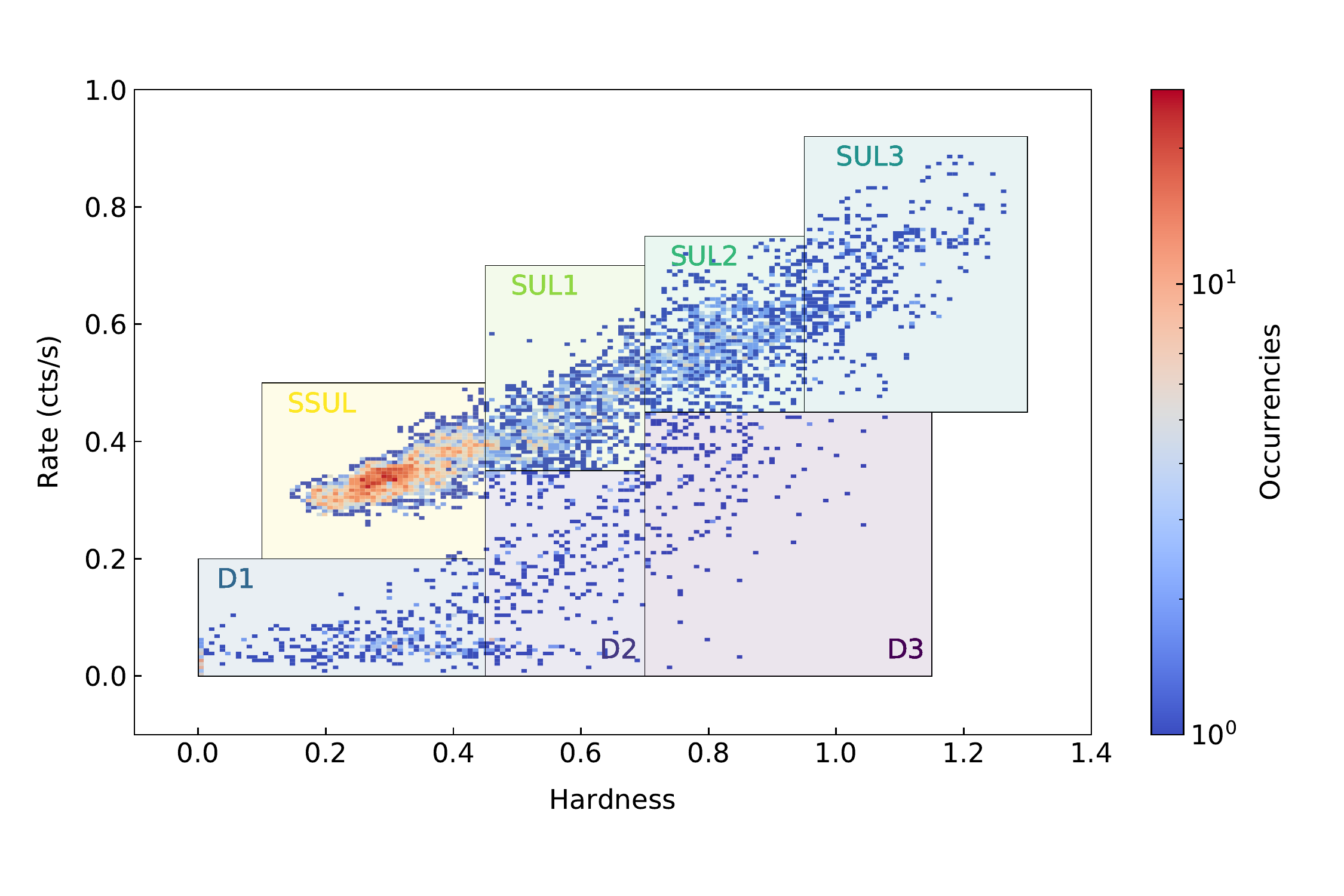}
\vspace{-0.5cm}
\caption{2D density plot of the hardness-intensity diagram from the 
whole set of \xmm\ observations examined in this work. Right bar show the colour gradient in log scale. Boxes and labels show the
regions used for the spectral products extraction.}
\label{fig:hid}
\end{figure*}

To investigate in detail the spectral changes associated with the flux and hardness variations, we performed a hardness-intensity diagram (HID) resolved spectral analysis. To define the soft and hard energy bands, we took the median photon energy of the full time-averaged spectrum, approximately 0.9 keV, the energy at which there is also the strongest spectral curvature \citepalias[see, e.g.,][]{pinto21}. 
We extracted energy-selected, background-subtracted, corrected light curves in the 0.2-0.9 keV and 0.9-10.0 keV energy bands with 100 s bin size and computed the hardness ratio (HR; defined as the hard-to-soft count rate ratio). However, due to the low statistics, the HR suffers from a high Poisson noise. To mitigate this noisy HR, we fitted each ObsID light curve with an interpolating smoothing function \footnote{We adopted the Savitzky-Golay filter present in the \textit{SciPy} module (\url{https://docs.scipy.org/doc/scipy/reference/generated/scipy.signal.savgol_filter.html}) with 15--21 typical windows depending on the variability associated to each ObsID}. We then computed the HR curve based on the hard and soft interpolating function values. In Fig.~\ref{fig:hid} we show the resulting HID as a 2-D density histogram, where colours indicate the point density, and, consequently, the exposure time.

The total HID shows two distinct branches, which We shall refer to as the \emph{normal} and \emph{dipping} branches. To study the spectral changes occurring in the different zones of the HID, but still working with high signal-to-noise ratio (SNR) spectra, we selected a total of seven regions onto the HID from which we extracted the relative good-time intervals (GTIs). The limits of the boundary regions were chosen to ensure enough statistics per spectrum as follows: HR $<0.45$, 0.45 $\leq$ HR $< 0.70 $, HR $\geq$ 0.70. Then, since in the normal branch the HR extends up to about 1.3 and the statistics is higher, we divided the third region of this branch in two parts at HR\,=\,0.95. These regions are shown in Fig.~\ref{fig:hid}. 
Observations with high variability and presence of many dips tend to populate the dipping branch, and, preferentially, the upper part of the normal branch as shown for ObsIDs 401, 501 and 601 (see Fig.~\ref{fig:lcurves}, where the light curve points are coloured according to the colour scheme of the regions of Fig.~\ref{fig:hid}). Observations characterised by low variability occupy a compact region of the HID in the bottom part of the normal branch (see e.g. ObsID 301, 701 and 801, Fig.~\ref{fig:lcurves}). ObsID 201 is in-between the two classes, as its points populate both the dipping branch and the lower part of the normal branch, due to the presence of just one, moderately-long, dip at the beginning of the observation, after which the source goes rapidly in the middle and, at the end of the observation, in the bottom of the normal branch. Both branches show a positive correlation between count rate and hardness, thus higher observed rates correspond to harder spectra. 

We will label SSUL the supersoft state region at the bottom of the normal branch; from SUL1 to SUL3 the regions of the upper part of the normal branch; from D3 to D1, the regions from the upper to the lower part of the dipping branch. From these HID selections, we derived the GTIs to filter both source and background spectra for all the EPIC instruments in each ObsID. After having verified the full compatibility of the MOS1 and MOS2 spectral products, we produced stacked spectral products (source, background and response matrices) using the \texttt{epicscombine} tool and refer to these combined products as the MOS12 products. 
PN and MOS12 spectra belonging to the same HID region from any ObsId were finally added using \textsc{addascaspec}.

We re-binned MOS12 and PN spectra with the \textsc{specgroup} tool, adopting a minimum SNR of three, requiring no more than three energy channels per resolution element, and a minimum number of 25 counts per channel to use the $\chi^2$ statistics. We performed spectral fits in the 0.6--10 keV and 0.3--10 keV bands for PN and MOS12 spectra, respectively. We report the total exposure times and average count rates for the intensity-HID resolved PN and MOS spectra in Table~\ref{tab:expo_rates}. 

\begin{table}
    \centering
    \begin{tabular}{cc|rr |rr } \hline
             & &         \multicolumn{2}{c}{Exposure times} & \multicolumn{2}{c}{Count rates} \\
             & &         \multicolumn{2}{c}{10$^4$ s} & \multicolumn{2}{c}{10$^{-2}$ counts s$^{-1}$} \\
          & State    &           PN & MOS12 & PN & MOS12 \\  
              \hline 
      \multirow{4}{*}{\it{Normal branch}}
         &SSUL    &  29.3   & 33.9  & 19.7 & 10.0\\
         &SUL1    &  14.4  & 16.3  &  27.6 & 15.2 \\
         &SUL2    &  7.9   & 9.0   & 39.3 & 21.8 \\
         &SUL3    &  3.4   & 4.1   & 47.1 & 26.3 \\
         \hline
         \multirow{3}{*}{\it{Dipping branch}}
        & D3    &  1.0  & 1.2 & 20.8 &  12.5 \\
        & D2    &  2.4  & 2.7  & 8.6  &  4.4 \\
        & D1   &   3.8  & 4.3  & 3.2  &  1.7 \\
         \hline
    \end{tabular}
    \caption{Total exposures and background-subtracted average count rates 
    (0.6-10 keV and 0.3-10.0 keV energy bands for PN and MOS12, respectively) for the PN and MOS12 HID-selected spectra.}
    \label{tab:expo_rates}
\end{table}
{\centering
\begin{figure*}
\begin{tabular}{cc}
\tabularnewline
\vspace{-0.1cm}
\includegraphics[width=\columnwidth, height=0.62\columnwidth]{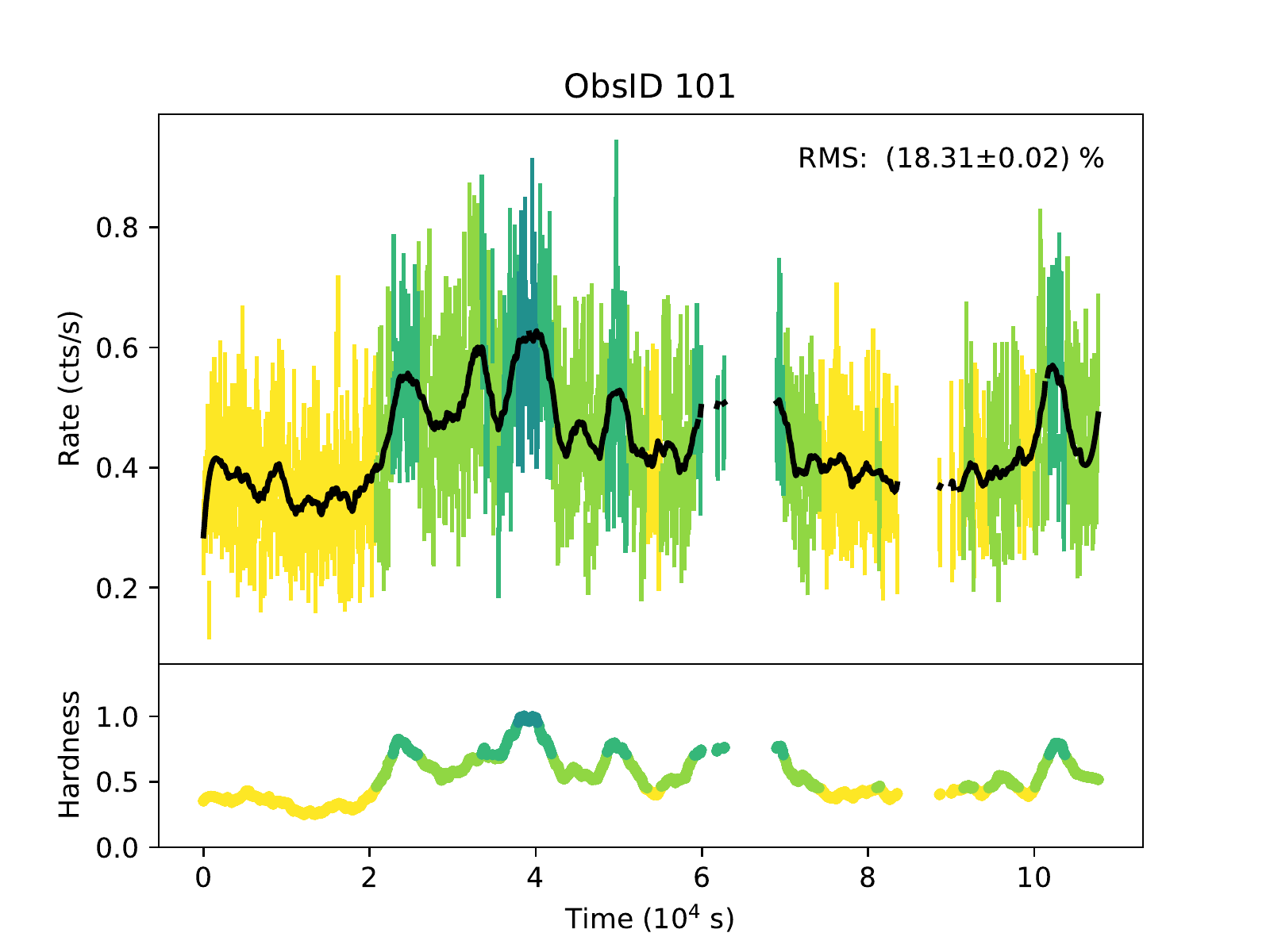} &
\includegraphics[width=\columnwidth, height=0.62\columnwidth]{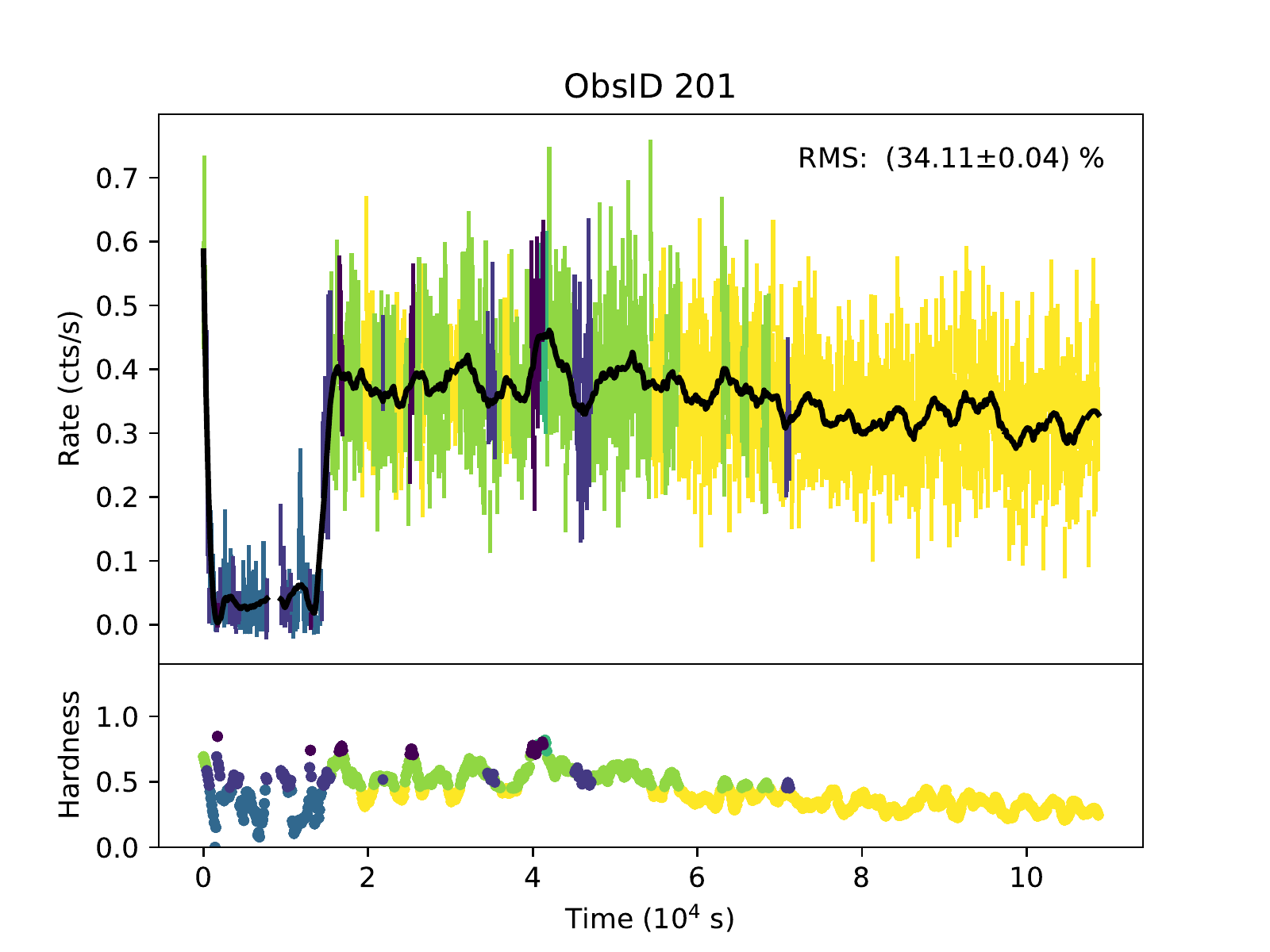} \\
\vspace{-0.1cm}

\includegraphics[width=\columnwidth, height=0.62\columnwidth]{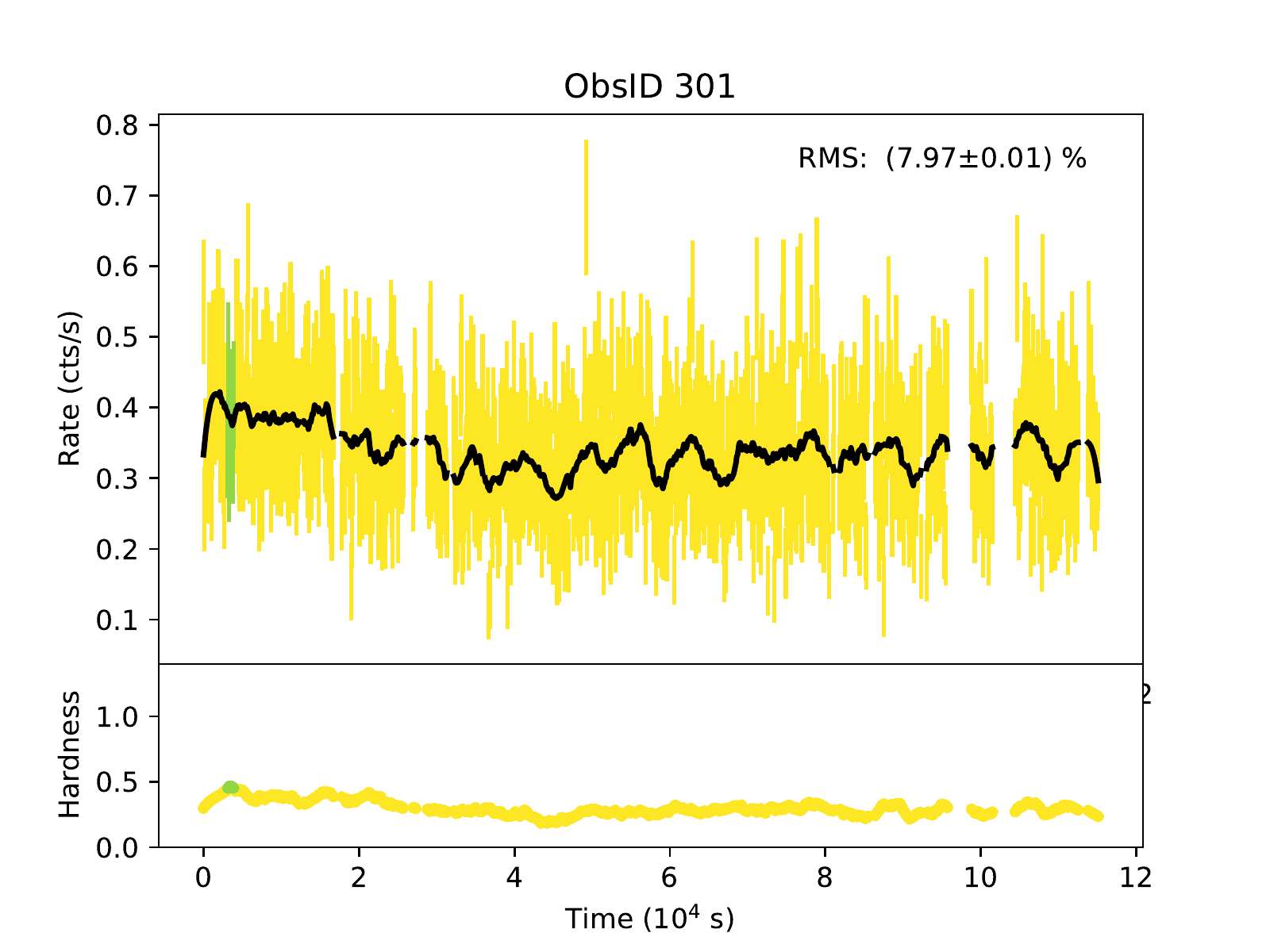} & 
\includegraphics[width=\columnwidth, height=0.62\columnwidth]{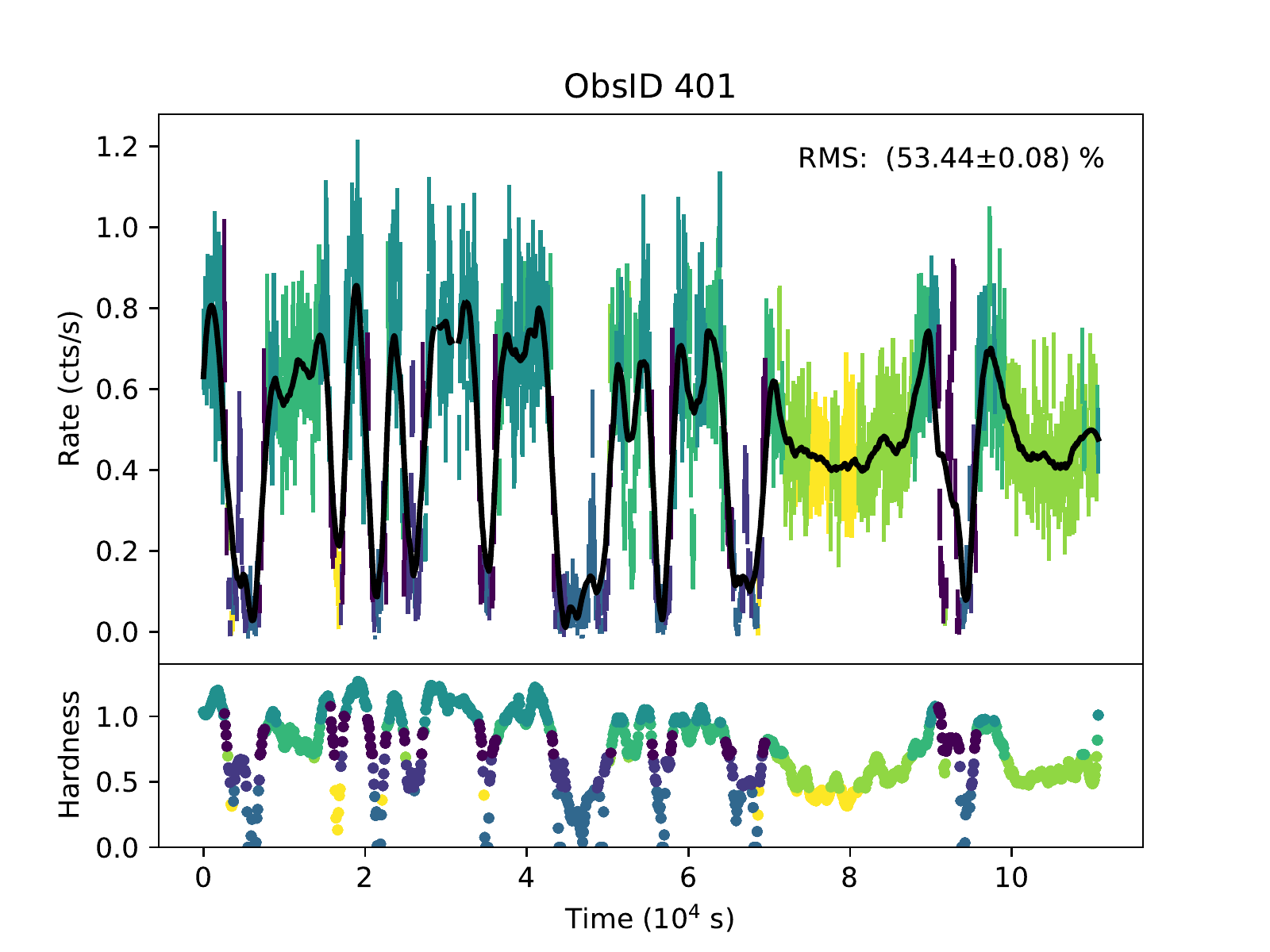} \\
\vspace{-0.1cm}
\includegraphics[width=\columnwidth, height=0.62\columnwidth]{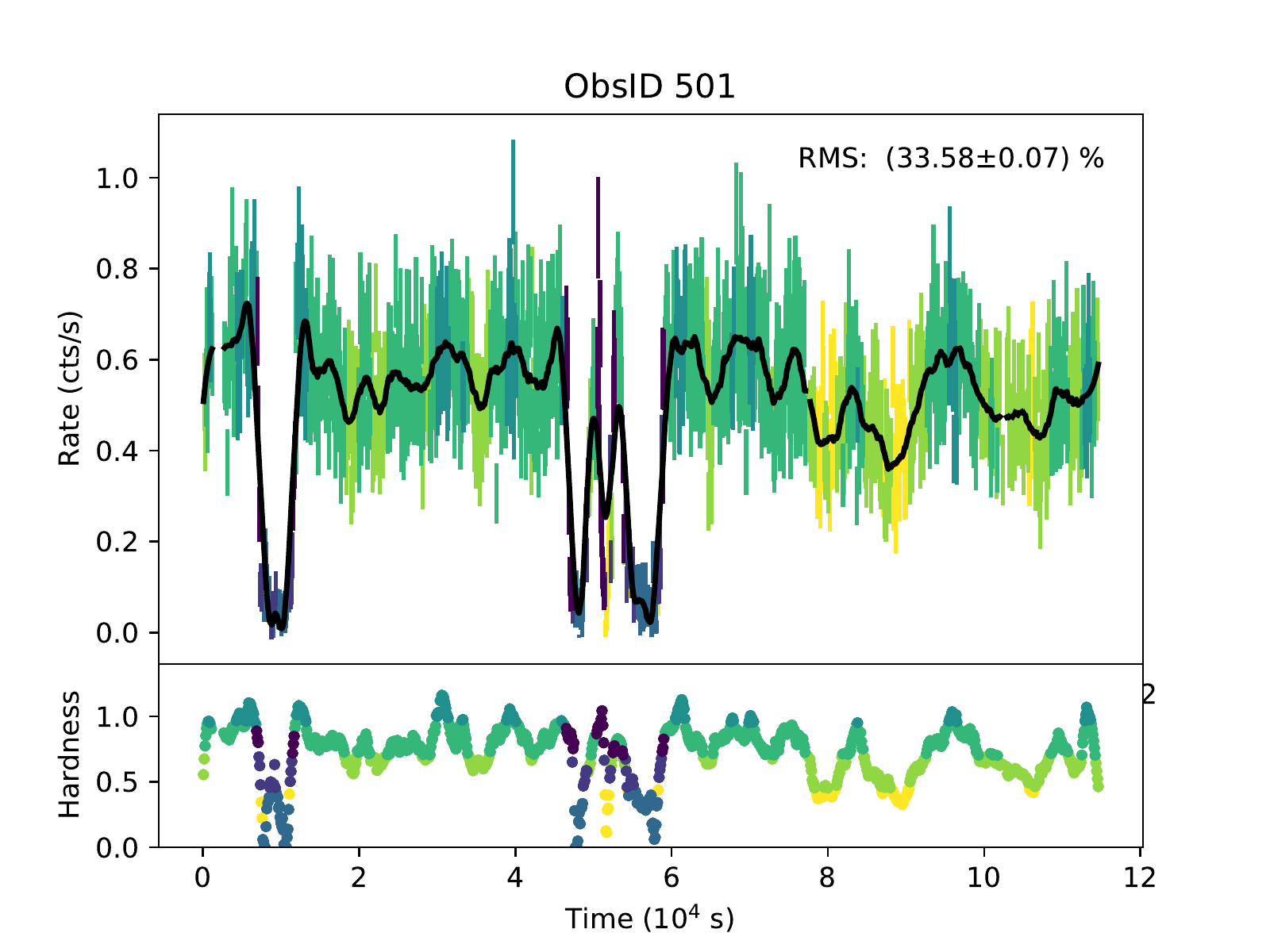}&  
\includegraphics[width=\columnwidth, height=0.62\columnwidth]{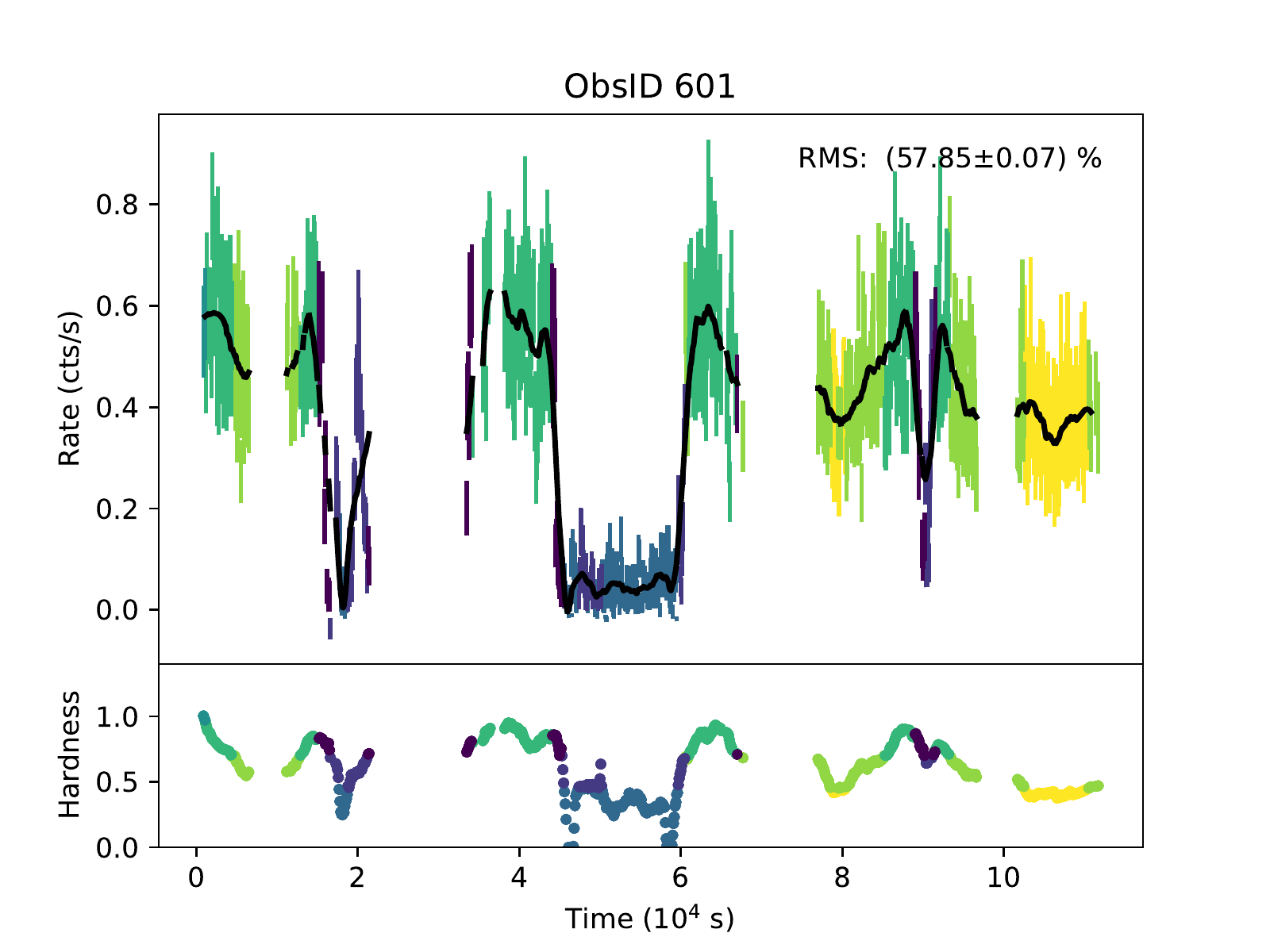} \\
\vspace{-0.1cm}
\includegraphics[width=\columnwidth, height=0.62\columnwidth]{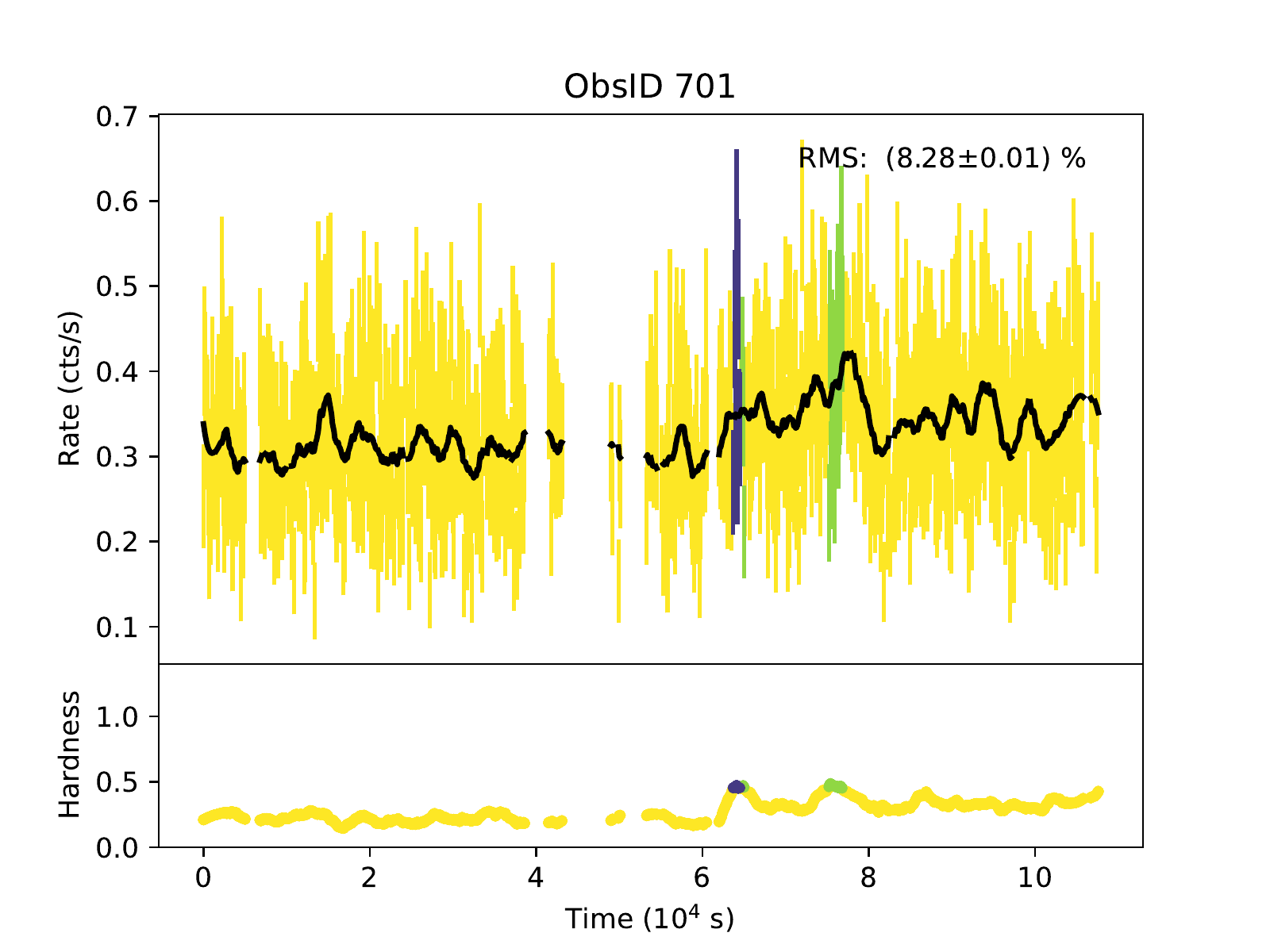} &  
\includegraphics[width=\columnwidth, height=0.62\columnwidth]{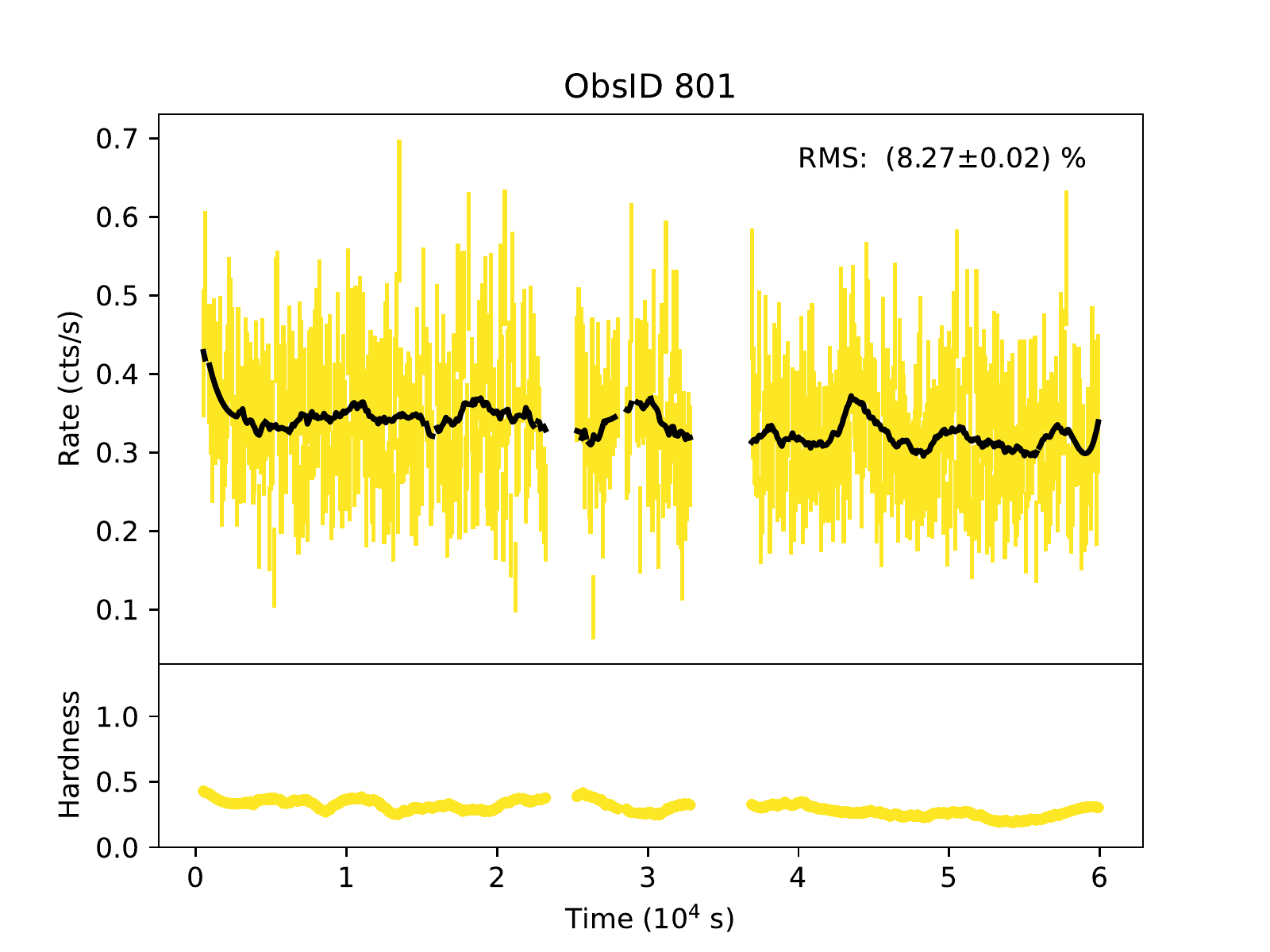} 
\end{tabular}
\caption{EPIC/PN light curves of each ObsID (time bin-size 100s). 
Over-plotted the best-fitting smoothing curve (continuous black line). Lower panels show the 
corresponding hardness (0.9-10 keV / 0.2-0.0.9 keV) curve. Colours identify the good-time intervals used
for spectral extraction according to the colour scheme shown in Fig.\ref{fig:hid}.}
\label{fig:lcurves}
\end{figure*}
}

\section{Fit procedure and results} \label{sect:hidselectedspectra}

The continuum emission of this source has been modelled in the past using different combinations of soft thermal component and a power-law for the hard component \citep[][]{jin11, feng16}. The continuum time-averaged spectrum of all these observations has been modelled by \citetalias{pinto21} with a combination of three black-bodies. For our HID-selected spectra, we found that the combination of a soft black-body and a  multi-colour disk black-body component (\texttt{diskbb} in \textsc{xspec}) resulted in satisfactorily fits without the need of an additional third component \citep[see][for a similar approach for the similar source NGC 55 ULX-1]{pintore15}. We caution the reader that referring to this disk component flux and temperature in what follows, we do not interpret it physically associated with emission from a standard sub-Eddington disk. The spectral disk shape serves to fit all the internal hard X-ray reprocessed emission, whose \textit{primary shape} is not accessible to us.

This continuum is absorbed by an equivalent hydrogen column density $N_{\textrm H}$ modelled with the \texttt{tbabs} component, with element abundances from \citet{wilms00} and atom cross-sections from \citet{verner96}. 
For all fits the $N_{\textrm H}$ parameter was fixed to the best-fit value of the time-averaged (over all the observations) spectrum (1.7\,$\times$10$^{21}$ cm$^{-2}$) as we noted this parameter did not significantly vary for the different spectra.

The correct modelling of the emission/absorption features is not trivial. 
To limit the model parameter degeneracy due to the low energy resolution of the EPIC in the soft band, the apparent broadening of the line profiles, the shifts in the line positions in correlation with the flux state, we took advantage of the time-averaged high-resolution RGS analysis reported in \citetalias{pinto21}.  
From the Gaussian line scan, we retrieved the energy of the most significant emission/absorption lines.
We interpret these lines as generated by two different plasma environments. For the emission line pattern, we fixed the normalisations and the centroids of all the lines to the RGS best-fit results (see Table \ref{tab:lines}) and left in the fit procedure a common multiplicative flux constant ($C_{\rm em}$) free to vary for the whole set of lines. This thin plasma, is parametrised as a co-moving single region where lines share a common width ($\sigma_{\textrm{em}}$).  Our set of fitted lines does not comprehend all the possible transitions involved in a physical plasma model as shown in \citetalias{pinto21}. Although in the RGS all the lines are resolved with a few eV width, the EPIC spectra strongly requires a width larger than the spectral resolution of the PN and MOS CCDs ($\sim$\,100 eV and 60 eV, full-width half maximum at 1 keV, respectively). In our choice this broadening accounts for the vaster pattern of close-by blended lines and a possible state-dependent physical broadening. As a necessary further check, we also re-fitted our spectra by leaving the line widths of the single detected lines free to vary in a closed neighbourhood (around 0.1 keV) of our best-fit parameters, and we did not note any statistically significant difference from the common best-fit values. 

\begin{table}
    \centering
    \begin{tabular}{rrrr} \hline
    &$E_{\rm fix}$ & $E_{\rm rest}$ (Id.) & $N_{\rm fix}$ \\
    &keV & keV & \tiny{$10^{-6}$ ph cm$^{-1}$ s$^{-1}$} \\ \hline
    &\multicolumn{3}{c}{Emission lines} \\ \hline
    \texttt{zgau$_1$} & 0.5920 & O\,{\scriptsize{VII-VIII}} & 5.8\\
    \texttt{zgau$_2$} & 0.8865 & Ne\,{\scriptsize{IX}}-Fe\,{\scriptsize{XVIII}} & 8.5 \\
    \texttt{zgau$_3$} & 0.9766 & Ne\,{\scriptsize{X}}-Fe\,{\scriptsize{XX-XXI}} & 6.6 \\
    \texttt{zgau$_4$} & 1.5540 & Mg\,{\scriptsize{XI-XII}}                      & 0.33 \\ \hline
    &\multicolumn{3}{c}{Absorption lines} \\ \hline
    \texttt{zgau$_5$} & 0.7483 & O\,{\scriptsize{VIII}} & -0.544 \\
    \texttt{zgau$_6$} & 1.1813 & Ne\,{\scriptsize{X}}-Fe\,{\scriptsize{XXI}}  & -2.73 \\
    \texttt{zgau$_7$} & 1.2845 & Fe\,{\scriptsize{XXIII-XXIV}}                & -2.34 \\
         \hline
    \end{tabular}
    \caption{Emission and absorption line patterns as estimated from the EPIC-RGS time-averaged spectrum \citepalias[see][]{pinto21}. Energy positions ($E_{\rm fix}$) and line normalisations ($N_{\rm fix}$) are frozen in the fits. Absorption/emission lines have tied line widths ($\sigma_{\rm em}$ and $\sigma_{abs}$) and a common multiplicative constant factor ($C_{\rm abs}$ and $C_{\rm abs}$), left free to vary in the fits (parameters reported in Table \ref{tab:fitresults}).}
    \label{tab:lines}
\end{table}

We observed significant shifts in the line energies for the different HID-resolved spectra \citepalias[similarly to what already shown in the spectra averaged over the single ObsIDs, see][]{pinto21}. To consistently evaluate these shifts we assume that the relative line pattern remains unchanged for the different spectra and let a common shift parameter free to vary. For this fitting purpose, we modelled the lines using the \texttt{zgauss} component, and the red-shift parameter $z_{\rm em}$ is free to vary. 
This choice proved \textit{a posteriori} a satisfactorily way to flatten the residuals at the different line energies without the risk of increasing the parameters degeneracy.

The same fitting procedure is adopted for the complex of the absorption lines; in particular, we introduced in analogy with the emission line pattern a similar set of parameters: $C_{\rm abs}$, $\sigma_{\rm abs}$, and $z_{\rm abs}$ for the relative multiplicative absorption normalisation constant, common absorption line width, and relative absorption line red-shift, respectively. In the fitting process, if we obtained only an upper limit to the $C_{\rm abs}$ value (the 90\% c.l. value is reported in Table \ref{tab:fitresults}), we set this parameter to 0 (freezing also the $\sigma_{\rm abs}$ and $z_{\rm abs}$ parameters) for the computation of best-fitting values and uncertainties of all other parameters.  Finally, a  normalisation constant between the PN and the MOS12 spectra is left free to vary to account for any residual flux inter-calibration mismatch. The constant is frozen to 1 for the PN spectrum, and free to vary for the MOS12 model ($C_{\rm mos12}$ parameter). Our model in \texttt{xspec} reads as: \texttt{C$_{\rm mos12}$ $\times$ tbabs $\times$ (bbody+diskbb+C$_{\rm em}$ $\times$(zgau$_{1}$ + zgau$_{2}$ + zgau$_{3}$ + zgau$_{4}$) + C$_{\rm abs}$ $\times$ (zgau$_{5}$ + zgau$_{6}$ + zgau$_{7}$))}.

We emphasise that, despite our best efforts the modelling described above is still limited by quality of the available data, i.e. counting statistics and energy resolution of the EPIC cameras. Nevertheless, our results appear consistent with the full time-averaged RGS results \citepalias{pinto21}, so we are confident that they are reasonably robust. Future observations with instruments with higher energy resolution such as \textit{XRISM} \citep{2020arXiv200304962X} or \textit{ATHENA} \citep{2020AN....341..224B} could be beneficial for firmly constraining this variable emission/absorption line landscape.

\begin{table*}
\caption{Best-fit results. Unabsorbed luminosity calculated in the 0.01--10 keV range with respect to the \emph{EPIC/PN} spectrum,  
assuming isotropy and 3.4 Mpc distance. For $R_{\textrm{disk}}$ we assumed a disk inclination of 60\degmark. $C_{\textrm{em}}$ and $C_{\rm abs}$ are multiplicative model normalisation constants for the optically thin/thick line emitting/absorbing line patterns. $L_{\rm abs}$ is the difference in luminosity between the best-fit model and a model with $C_{\rm abs}$ set to zero. Errors are given at 1 $\sigma$ c.l. }

\begin{tabular}{l| lll lll lll}
\multicolumn{10}{c}{Continuum}\\
\hline
Spectrum  &
$kT_{\rm bb}$    & $R_{\rm bb}$    & $L_{\rm bb}$  &
$kT_{\rm disk}$  & $R_{\rm disk}$  & $L_{\rm disk}$  &
 $L_{\rm tot}$  & $C_{\rm mos12}$   \\
\hline
    &  
eV  &   10$^3$ km  & 10$^{38}$ erg s$^{-1}$  & 
eV  &   10$^2$ km  & 10$^{38}$ erg s$^{-1}$  & 
    10$^{38}$ erg s$^{-1}$ & \\
 \hline
SSUL & 147$\pm$3 & 4.73$\pm$0.19 & 12.9$\pm$1.1  & 510$\pm$40  & 1.0$\pm$0.3    & 0.88$\pm$0.09& 15.01$\pm$0.22 &1.032$\pm$0.007 \\  
SUL1 & 160$\pm$4 & 3.60$\pm$0.13 & 11.5$\pm$0.5  & 500$\pm$20  & 2.1$\pm$0.3    & 3.3$\pm$0.4& 16.6$\pm$0.3 &1.041$\pm$0.009   \\ 
SUL2 & 188$\pm$3 & 2.96$\pm$0.08 & 13.3$\pm$0.5  & 610$\pm$20  & 1.73$\pm$ 0.17 & 5.1$\pm$0.4  & 20.3$\pm$0.3 & 1.019$\pm$0.010 \\
SUL3 & 200$\pm$4 & 2.68$\pm$0.09 & 14.3$\pm$1.0  & 640$\pm$30  & 1.9$\pm$0.3    & 7.5$\pm$0.7    & 23.5$\pm$0.4 &1.002$\pm$0.012 \\
D3 & 173$\pm$6   & 2.63$\pm$0.25 & 7.5$\pm$0.9  & 640$\ddagger$  & 1.21$\pm$0.05  & 3.0$\pm$0.3 & 9.7$\pm$0.6 &1.09$\pm$0.04     \\  
D2 & 140$\pm$4   & 3.3$\pm$0.3   & 5.2$\pm$0.4  & 640$\ddagger$  & 0.73$\pm$0.03  & 1.1$\pm$0.1 & 6.7$\pm$0.5 & 1.08$\pm$0.04     \\%
D1 & 128$\pm$5   & 2.7$\pm$0.3   & 2.4$\pm$0.3  & 640$\ddagger$  & 0.37$\pm$0.03  & 0.30$\pm$0.05& 2.9$\pm$0.3 & 1.05$\pm$0.05   \\%
\hline
\multicolumn{9}{c}{}\\
\multicolumn{10}{c}{Emission/Absorption Line Pattern}\\
\hline
Spectrum &
$C_{\rm em}$   & $\sigma_{\rm em}$   & $z_{\rm em}$     &   $L_{\rm em}$ &
$C_{\rm abs}$  & $\sigma_{\rm abs}$  & $z_{\rm abs}$    &   $L_{\rm abs}$  & $\chi_{\nu}^{2}$/dof\\
\hline
&
& eV & 10$^{-2}$ & 10$^{38}$ erg s$^{-1}$ &
& eV & 10$^{-2}$ & 10$^{38}$ erg s$^{-1}$ & \\
\hline
SSUL & 4.8$\pm$0.4  & 87$\pm$5   & 5.6$\pm$0.5  & 1.70$\pm$0.08 &3.4$\pm$0.7 & 130$\pm$17 & 3.1$\pm$1.5 & -0.47 &150/94 \\ 
SUL1 & 4.9$\pm$0.6  & 110$\pm$8  & 1.9$\pm$0.7  & 1.9$\pm$0.2 &0.75$\pm$0.20 & 35$\pm$20 & -4.9$\pm$1.0 & -0.12 &155/117\\
SUL2 & 4.7$\pm$0.6  & 120$\pm$10 & -1.6$\pm$1.0 & 1.9$\pm$0.2 &<0.4$^{\dagger}$  & 50$\ddagger$  & 0$\ddagger$& $\ldots$ & 141/139\\  
SUL3 &  4.0$\pm$0.8 & 120$\pm$20 & -5.7$\pm$1.6 & 1.7$\pm$0.3 &<1.9$^{\dagger}$  & 50$\ddagger$  & 0$\ddagger$  & $\ldots$ &169/136  \\
D3   & 2.8$\pm$0.9  & 100$\pm$30 & -1.1$\pm$2.9 & 1.2$\pm$0.2 &<1.4$^{\dagger}$  & 50$\ddagger$ &  0$\ddagger$ &  $\ldots$ & 73/78\\
D2   & 1.2$\pm$0.4  & 50$\pm$30  & 6.5$\pm$1.9  & 0.43$\pm$0.11 &<3.0$^{\dagger}$  & 50$\ddagger$ & 0$\ddagger$ &  $\ldots$ & 61/58\\
D1   & 0.64$\pm$0.18& 50$\pm$30  & 8.2$\pm$2.2  & 0.21$\pm$0.06 &<0.14$^{\dagger}$ & 50$\ddagger$ & 0$\ddagger$  &  $\ldots$ & 38/41\\
\hline
\multicolumn{8}{l}{$\dagger$ 90\% upper limit calculate with $\sigma_{\rm abs}$=50 eV and $z_{\rm abs}$=0.  \quad
$\ddagger$ fixed parameter}\\
\end{tabular} 
\label{tab:fitresults}
\end{table*}

\begin{figure*}
\centering
\begin{tabular}{cc}
\vspace{-0.5cm}
\includegraphics[width=\columnwidth]{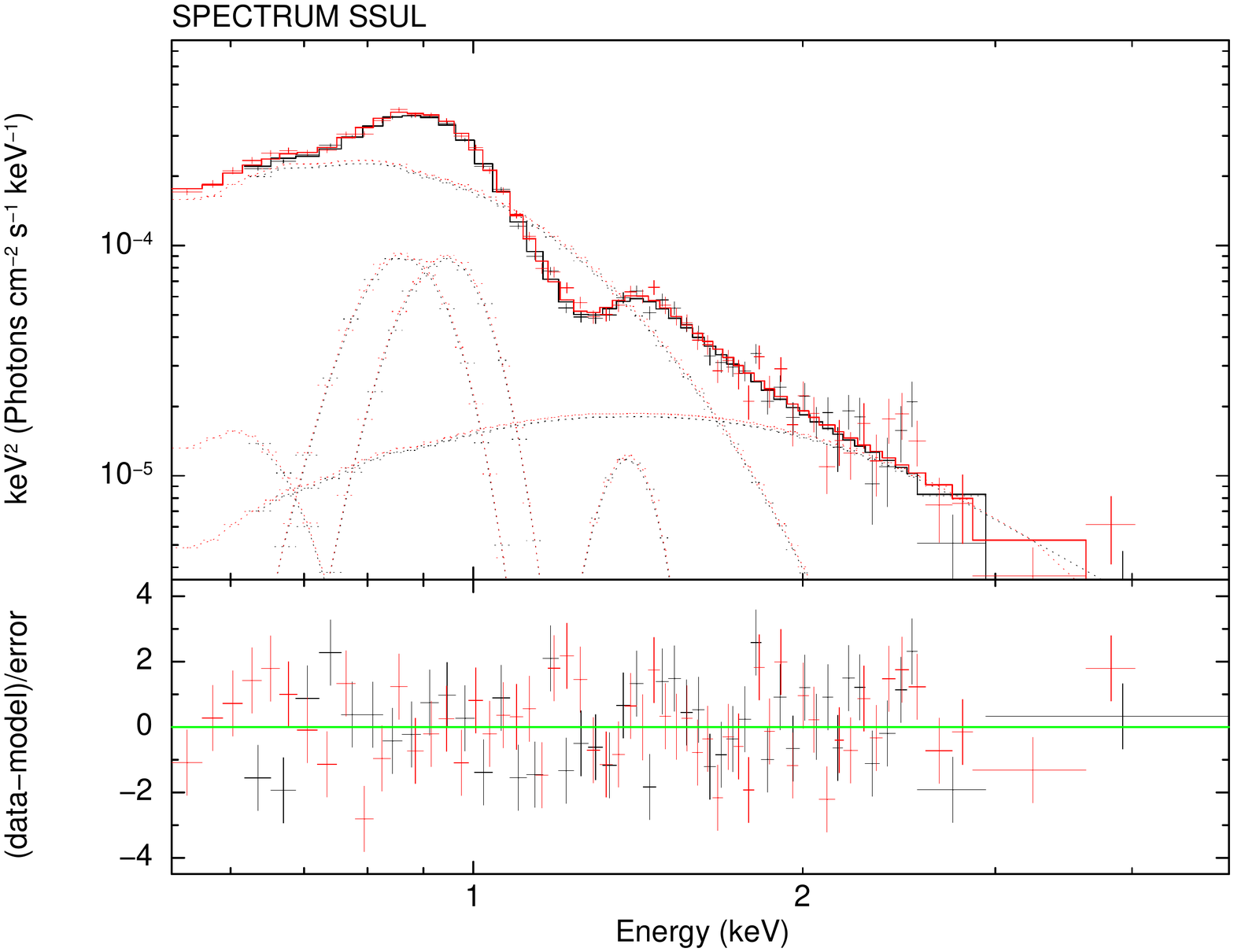}     & \includegraphics[width=\columnwidth]{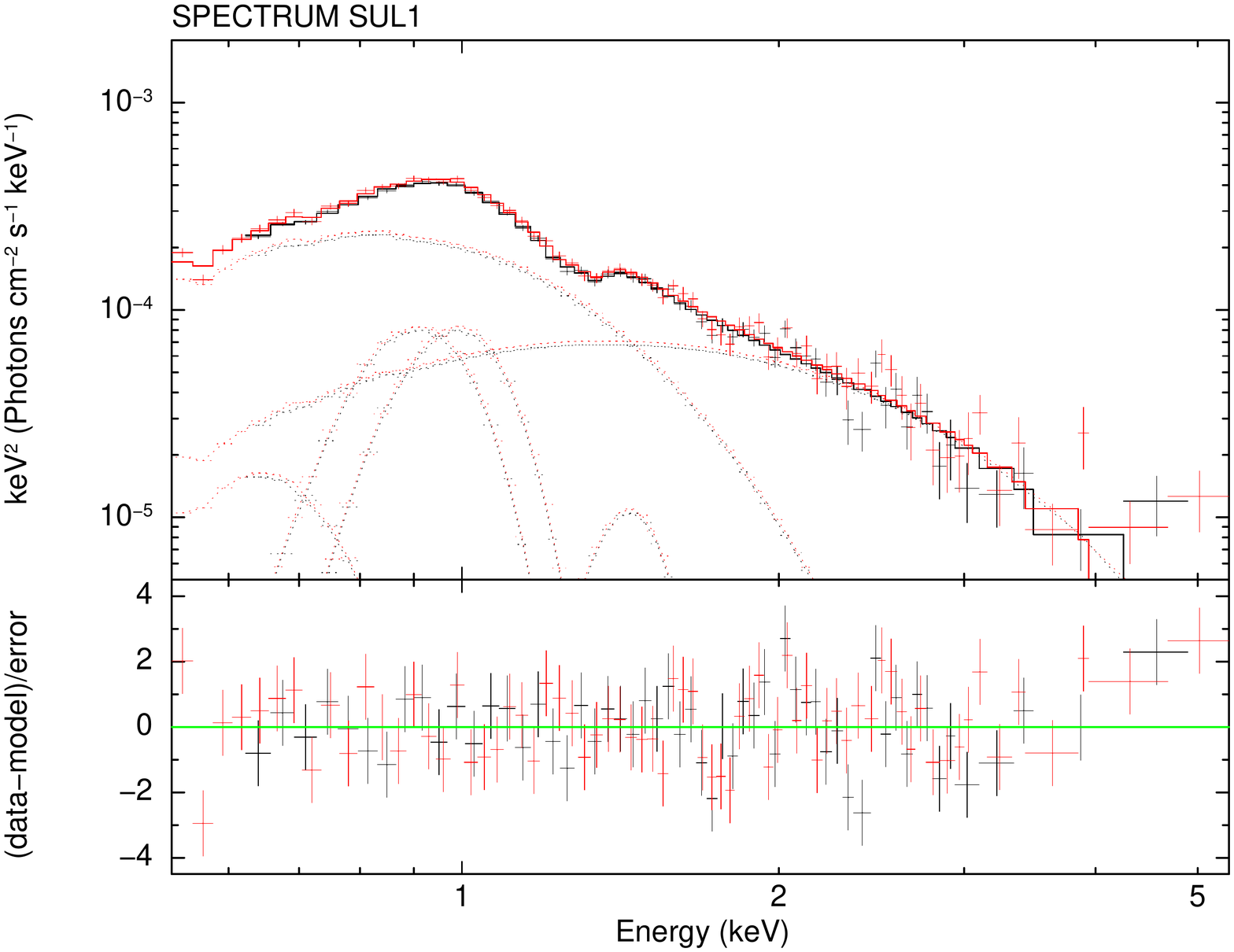} \\
\vspace{-0.5cm}
\includegraphics[width=\columnwidth]{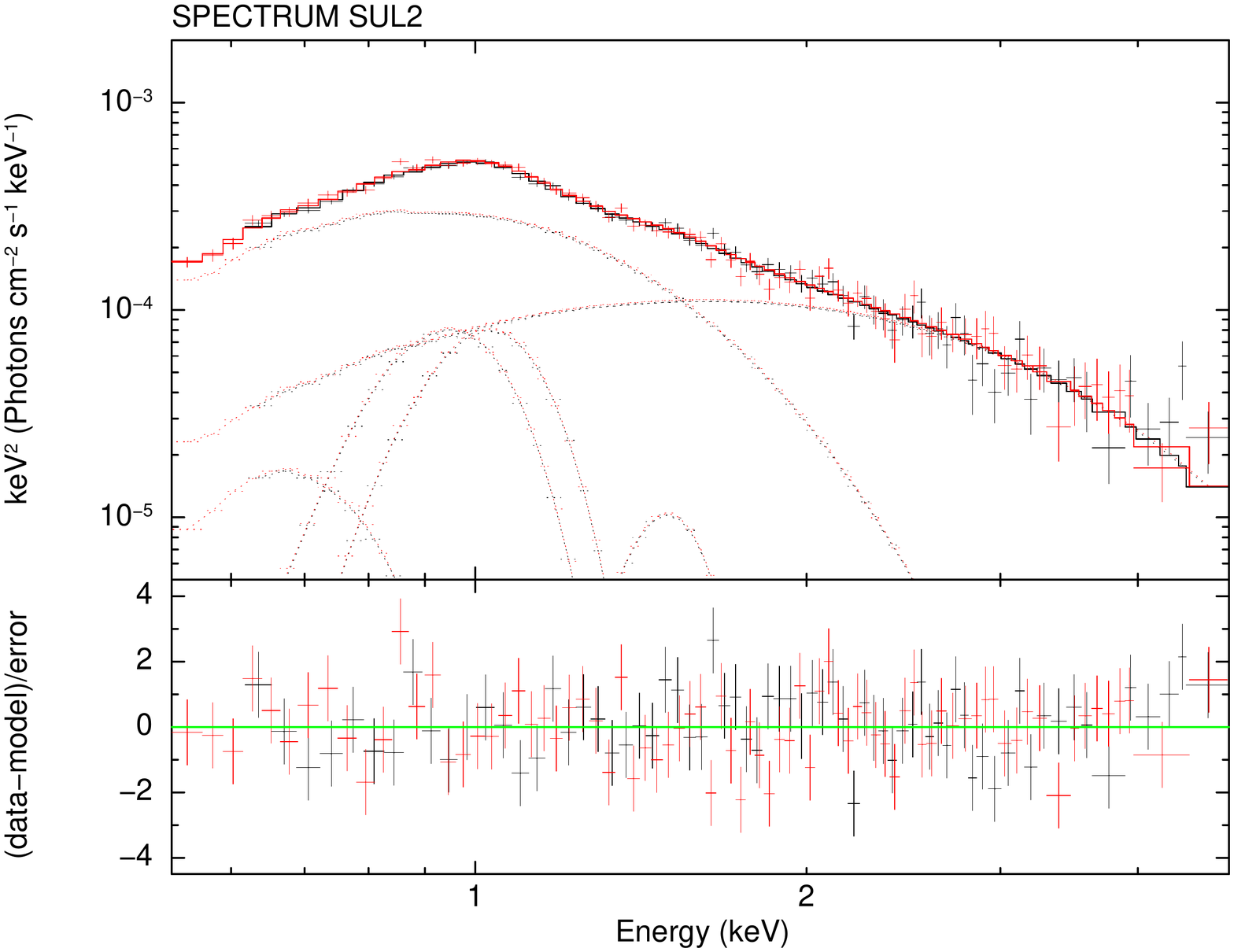}     & \includegraphics[width=\columnwidth]{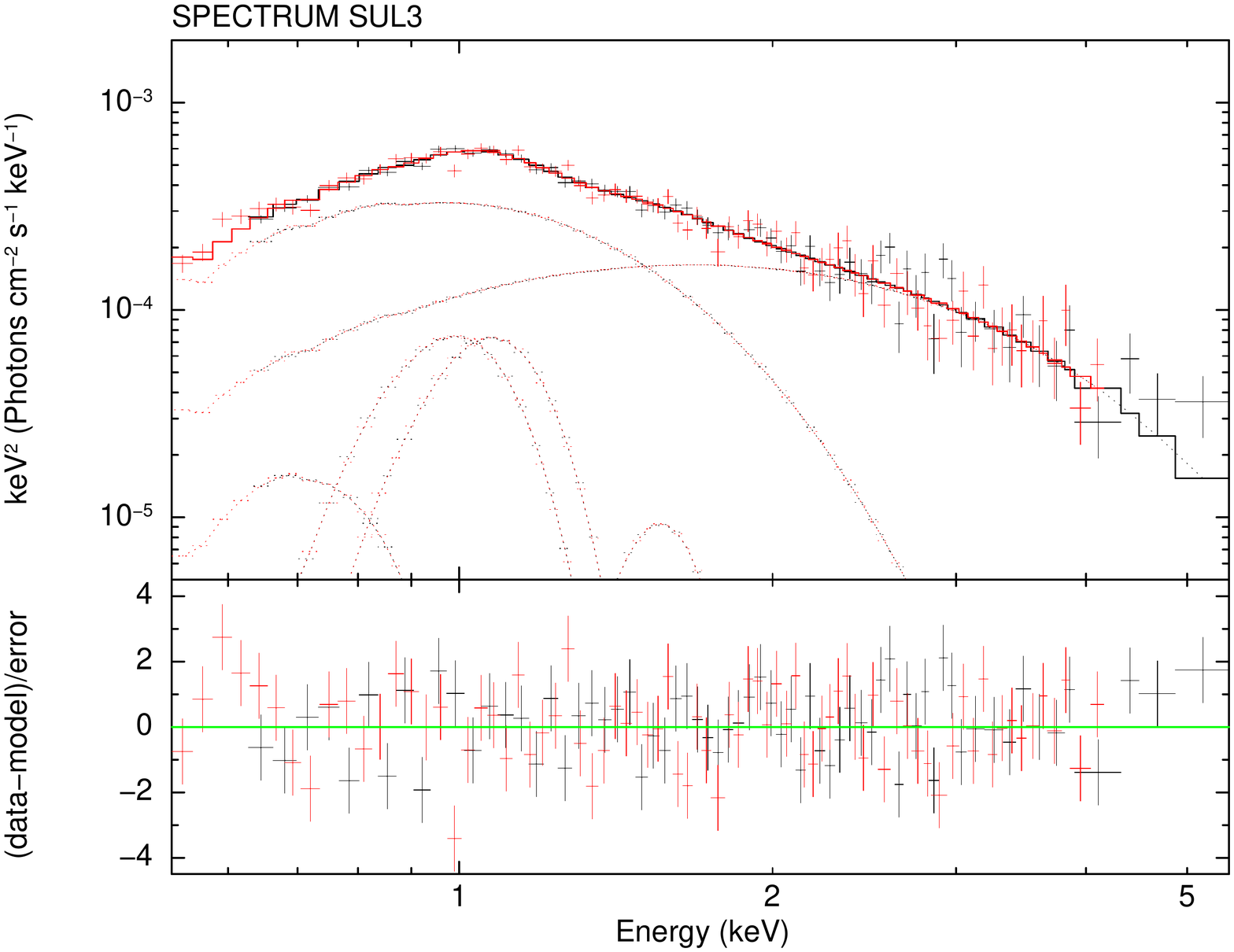} \\
\vspace{-0.5cm}
\includegraphics[width=\columnwidth]{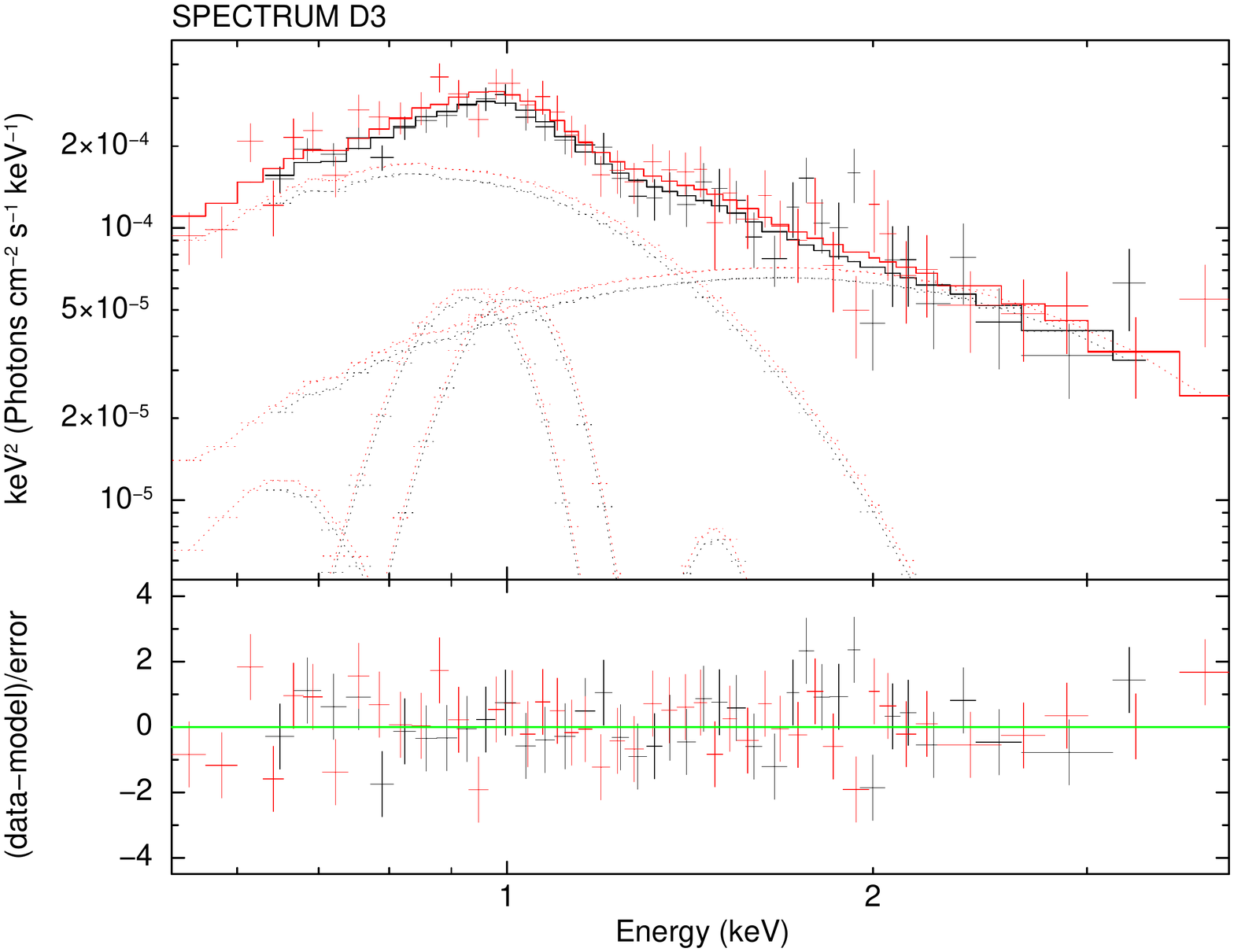} &  \includegraphics[width=\columnwidth]{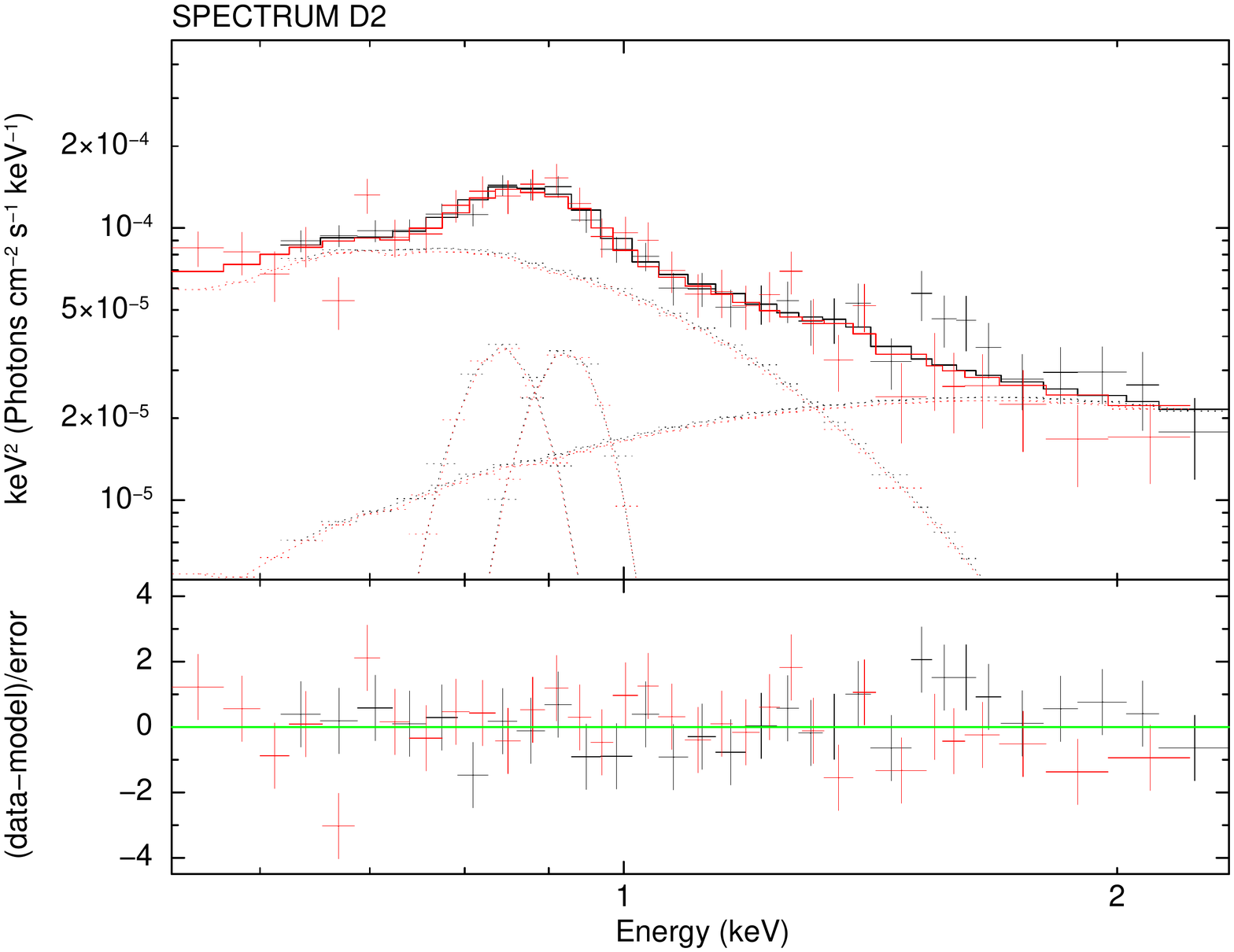} \\
\vspace{-0.3cm}
\includegraphics[width=\columnwidth]{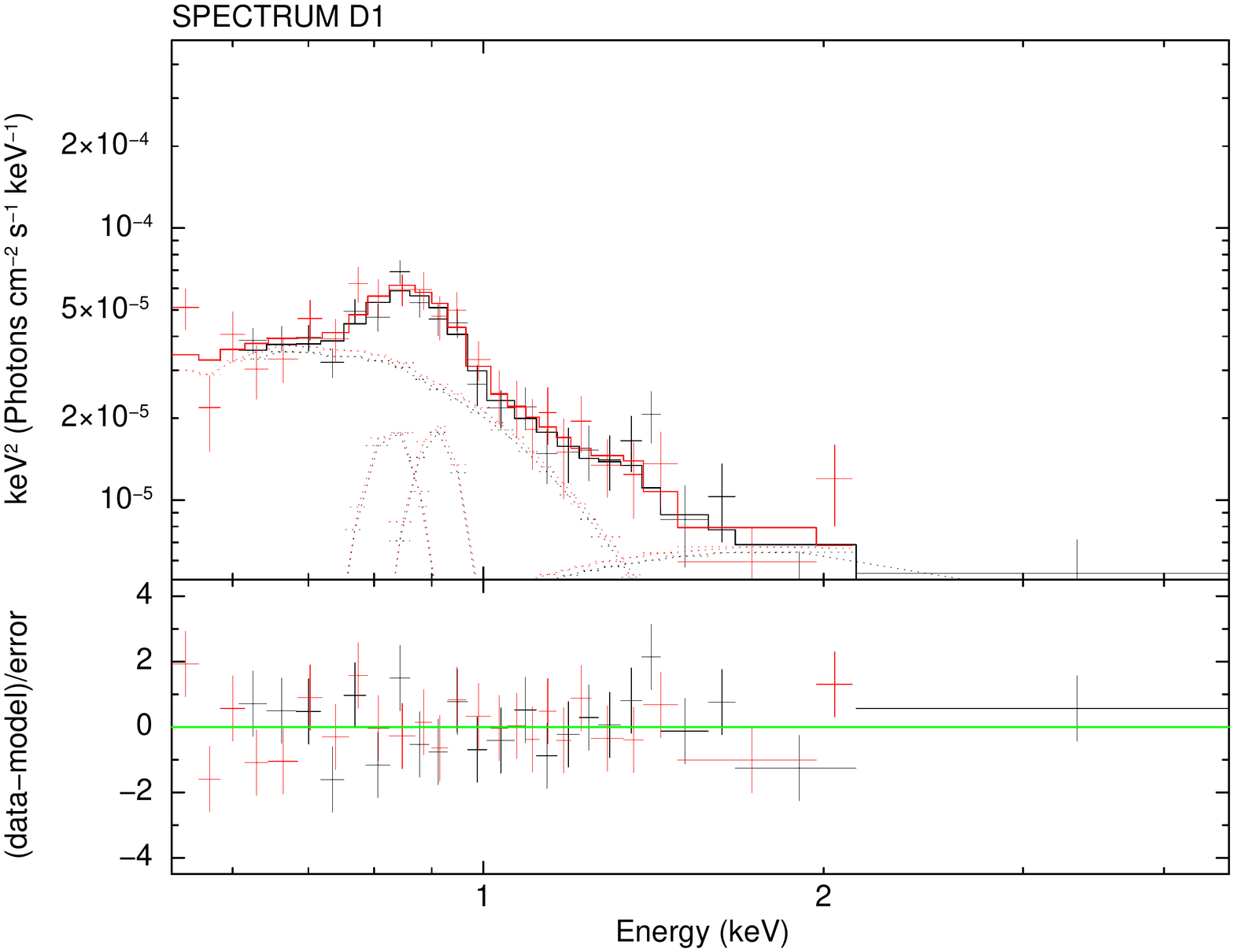} & \includegraphics[width=\columnwidth]{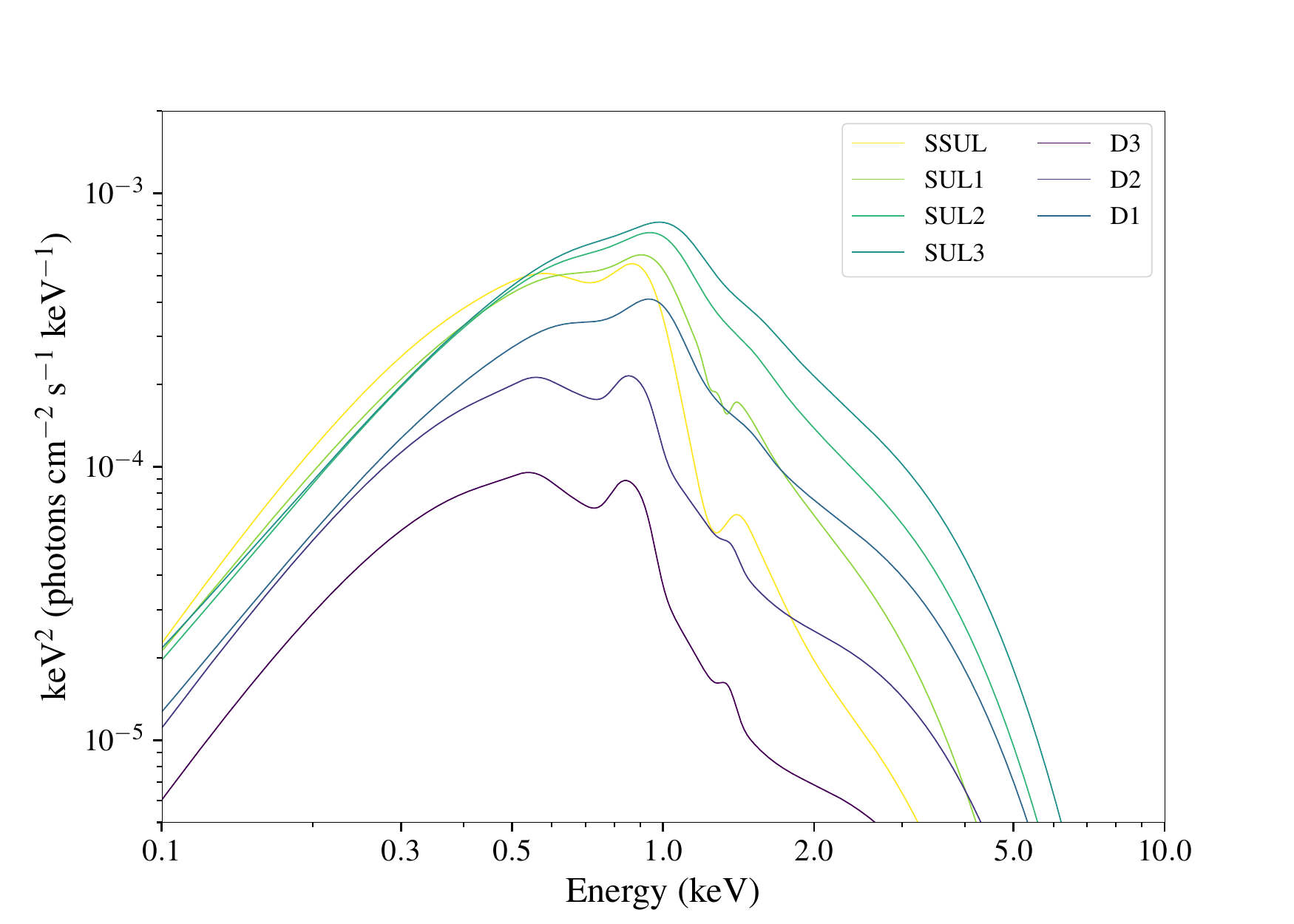}
\end{tabular}
\caption{Unfolded energy spectra and residuals in units of $\sigma$ for the all the HID-selected spectra (black and red data for the EPIC/PN and MOS12 data, respectively). Last figure comprehensively shows all the best-fitting unabsorbed models in the 0.1-10 keV range in the $E^2\times f(E)$ representation.}
\label{fig:spectra_normalbranch}
\end{figure*}

In Table~\ref{tab:fitresults} we report the final results of the fit for each selected spectrum. As it can be noted the final $\chi^2$ are not always formally acceptable (reduced $\chi^2$ $\gg$\,1), although the data overall appear reasonably well fitted (Fig.~\ref{fig:spectra_normalbranch}). The largest contributions to the final $\chi^2$ are due to unaccounted local features in energy bands which are not covered by the RGS spectral analysis (e.g. in the 2--3 keV range) or by unmodelled residuals in the low-energy part of the spectrum (0.5-0.7 keV) in some cases. 
In order to compare the spectral parameters without biases introduced by statistically poor local features (detection significance $\leq 3 \sigma$), we left these residuals unmodelled.
In Fig.~\ref{fig:spectra_normalbranch} we show the seven energy spectra and the unfolded unabsorbed best-fit models. 

The SSUL  state includes the low variability spectra (e.g. ObsID 301, 701 and 801). Because of the long integrated exposure, the stacked spectrum has a very high SNR. The absorption line pattern is clearly detected at a $\Delta \chi^2$\,=\,140 (statistical difference with respect a model where $C_{\rm abs}\,=\,0$).
The SUL1 spectrum gets a good fraction of the total exposure from ObsID 101, which shows no dips but a significant higher average count rate. 
The pattern of absorption lines is still detected, albeit at a lower significance with respect to SSUL, with $\Delta \chi^2$\,=\,22.
SUL2 and SUL3 spectra are populated by the most variable observations (e.g. ObsID 301, 401 and 501). We found no significant evidence for the absorption line pattern in any of these spectra.
The overall statistics for the dipping branch spectra is much lower than in the spectra extracted from the normal branch. For this reason, determination of all the parameters is subjected to larger uncertainties, particularly for the temperature and normalisation of the \textsc{diskbb} component.
As shown in Table~\ref{tab:fitresults}, the range of disk temperature values for normal branch spectra is 0.5-0.64 keV; at 90\% confidence level, we found a lower limit on the $kT_{\rm disk}$ value of 0.53 keV in the D3 spectrum, whereas in D1 and D2 any value of $kT_{\rm disk}$ in this range does not significantly change the $\chi^2$ value of the fit. 
We therefore freeze the $kT_{\rm disk}$ parameter in D1-D3 spectral fits to the best fitting value obtained for the SUL3 spectrum ($kT_{\rm disk}$\,=\,0.64 keV) because before any dip episode the most likely position of the source was in the SUL3 region. We note that setting this value to the lower limit ($kT_{\rm disk}$\,=\,0.5 keV) would raise the inner disk radius by a factor of $\sim$\,2, while the fluxes of all the spectral components and the other parameters would remain consistent within the statistical uncertainties reported in Table~\ref{tab:fitresults}. We found no statistical evidence for the presence of the absorption lines in any of these spectra, though the $C_{\rm abs}$ upper limits do not rule them out. 

\subsection{Spectral variability along the branches}
\src\ shows the highest luminosity ($\sim$2.3$\times 10^{39}$ \ergsec) in SUL3, whereas previous studies on this source reported on a maximum luminosity of about 10$^{40}$ \ergsec\ \citep{feng16}. We verified that such discrepancy is entirely due to our lower estimate of the absorption column.

\begin{figure}
\centering
\includegraphics[width=\columnwidth]{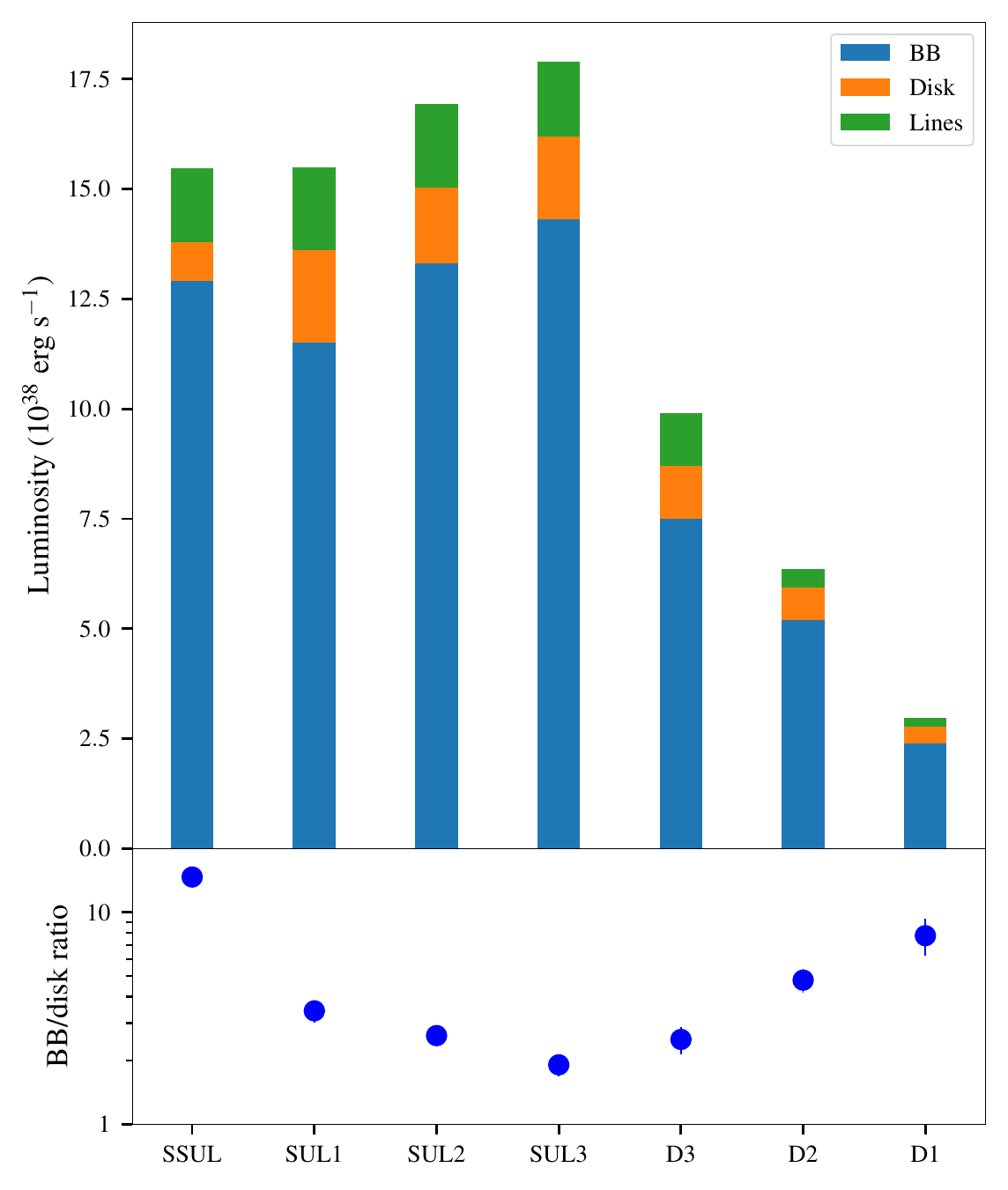}
\vspace{-0.5cm}
\caption{Unabsorbed luminosity (0.01-10 keV range) of the two continuum components as a function of the HID-selected regions.}
\label{fig:plots_fluxes}
\end{figure}

The bolometric luminosity decreases along the normal branch towards the SSUL of a factor of 1.5. This change in the luminosity is mainly driven by the \textsc{diskbb} component. Indeed, the fractional variation of the \textsc{bbody} component is only about 10\%, whereas the \textsc{diskbb} luminosity increases by a factor of 6 as shown in the upper panel of Fig.~\ref{fig:plots_fluxes}. The ratio of the bolometric luminosity (0.01-10 keV) between the black-body and the disk components in the SSUL state reaches a value $\sim$\,15, while in the SUL1-SUL3 this ratio ranges between 2 and 4 (lower panel of Fig.\,\ref{fig:plots_fluxes}). 
This is the most significant characteristic that distinguishes the SSUL state from all the other spectral states, as we did not find any evidence for \emph{intermediate} values.

\begin{figure}
\centering
\begin{tabular}{c}
\includegraphics[width=\columnwidth]{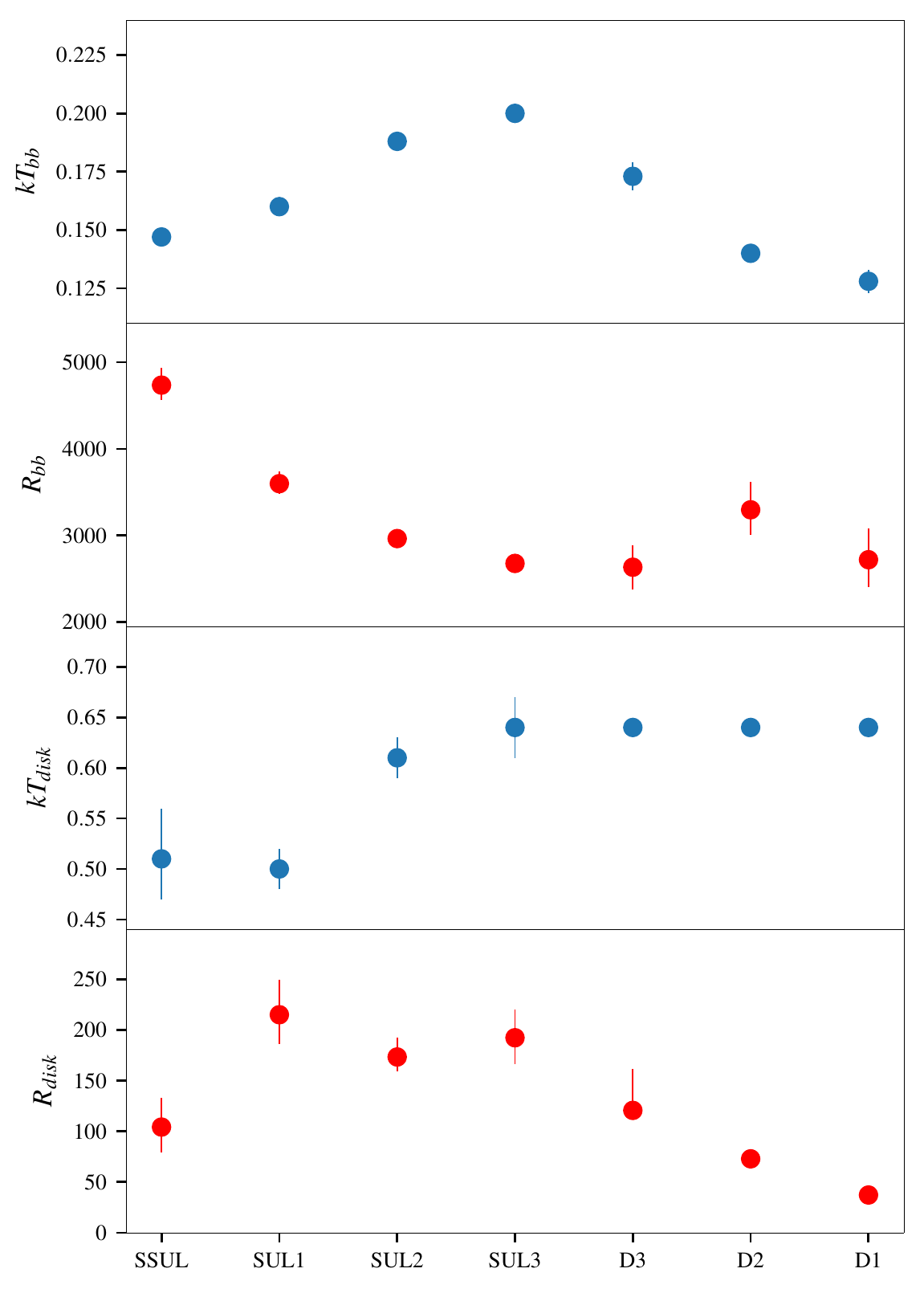} \\
\end{tabular}
\vspace{-0.3cm}
\caption{HID resolved continuum parameters evolution. From top to bottom panel: dependence of the black-body temperature, black-body radius, disk black-body temperature, and disk radius on the HID selected spectrum. Temperatures in units of keV, radii in units of km.
}
\label{fig:plots_disk}
\end{figure}

We show in Fig. \ref{fig:plots_disk} the basic parameter relations as a function of the source state for the two continuum components.
From SSUL along the normal branch we observe an increasing of the \textsc{bbody} temperature, from 0.15 to 0.2 keV, and 
a decreasing of the associated emitting radius. For the disk component, the SSUL$\rightarrow$SUL1 passage is mainly due to a geometrical change in the observable part of the system, i.e. more area from the \textsc{diskbb} component is visible, whereas the source shifts in the more variable part of the upper branch are due to (albeit small) changes in the emitting peak temperature of this component. 

The dipping branch is divided in three spectra: the D3 region characterises mainly the ingress/egress passages, the D2 region the intermediate dip flux and D1 the lowest flux observed in the dip. We observe that the bolometric luminosity rapidly decreases up to a factor of 10 compared to the normal branch, with a luminosity of $3 \times 10^{38}$ \ergsec\ in D1 (Table \ref{tab:fitresults}).
The softening observed along the dipping phase is due to both the decreasing of the black-body temperature (from 0.17 keV in D3 to 0.13 keV in D1) and the stronger attenuation of the disk emission of a factor of 10 from D3 to D1 compared to the black-body component which is of a factor of 3.

Finally, the emission lines pattern shows significant variability both in energy and width. 
These parameters appears to be strongly correlated with flux of the disk component.
On the other hand, the overall line flux is constant in the normal branch, while it decreases along the dipping. We point out that detailed investigation of the line emission/absorption pattern can only be constrained with the application of self-consistent photo- or collisionally-ionised models which requires combined EPIC and high-resolution spectra. However, this is beyond the scope of the present study.

\section{Search for periodic variability}

We performed a search for periodic signals in all 8 \xmm\ light curves of \src, following the procedure outlined in \citet{israel96}. Other than the standard procedure with a single power spectra density (PSD), we took into account period shifts, which may smooth out the signal in the single PSD, since such period variations are observed in most PULXs (Pulsating ULXs, see i.e. \citealt{israel16}). To this end, we corrected the time of arrival (ToA) of the source events, in each observation, for a grid of $\sim$\,10,000 points, by a factor of $-\frac{1}{2}\frac{\dot{P}}{P}\,t^2$, in the range 7$\times$10$^{-6}<|\frac{\dot{P}}{P}\,($s$^{-1})|<$1$\times$10$^{-11}$, where $\dot{P}$ is the first period derivative (see \citealt{rodriguez20} for details).

No significant coherent signals were detected; we calculated a 3$\sigma$ upper limit around 8\%--11\% for the pulsed fraction (PF) of signals in the data, PF here defined as the semi-amplitude of the signal sinusoid divided by the average count rate.
A marginal detection of a coherent signal at $\sim$\,23\,s was detected with PF $\sim$\,5\%, but which we were not able to confirm, as it was not present in any other data set after applying the corresponding corrections to the photons ToA, and the high number of PSD (about 90,000) statistically makes possible, in principle, the occurrence of such a signal by chance.

\section{Discussion} \label{sect:discussion}
In this paper we focused on the broadband spectral changes observed in a deep \xmm\ observational campaign of \src. Previous studies in this source already noted a spectral bimodality, where, on a typical \xmm\ observing window (around 100 ks), \src\ preferentially was found either in the so-called SSUL state, or in a relatively harder, more luminous state where dipping activity is also present (SUL state, \citealt{feng16}).
One interesting exception is found in ObsID 02, where we clearly spotted a full transition from a dip, followed by a period in the SUL state, and then to the SSUL on a time-scale of just a few hours. 
The source spent a 47\%, 41\% and 12\% of the observing time in the super-soft (SSUL), soft (SUL1-SUL3) and dip states (D1-D3), suggesting that transition from SSUL to SUL can happen once every few days. The fractions of low- and high-flux states is also consistent with previous \textit{Swift} monitoring campaign \citep[see][]{feng16}.
The SSUL state always occupies a well-defined and compact region in the HID and shows no significant spectral variability on time-scale as long as hundreds of ks. On the contrary, once the SUL state sets in, the source populates a larger HID area, creating a clear diagonal track. As shown in Fig.\ref{fig:hid}, the SSUL high density compactness might indicate a physical limit of the source or a tight geometric constraint. As dips occur, a lower branch is created beneath the normal one.

The source shows some preferred tracks in its movement along the HID. Dips occur as the source  is in the SUL regime, either from SUL2 or from SUL3 regions; ingress and egress times populate the D3 and D2 spectra, whereas the D1 spectrum comes from the time segments characterised by the lowest count rate (deep dip).  We noted that during the longest  dips, the source occasionally switched to 
harder flaring episodes in D2 state. Most dips shows the following pattern: SUL2/SUL3 $\rightarrow$ D3 $\rightarrow$ D2 $\rightarrow$ D1 $\rightarrow$ D2 $\rightarrow$ D3 $\rightarrow$ SUL2/SUL3. 
The passage from the SSUL to SUL state is only occasionally observed.
We selected seven regions on the HID and extracted the corresponding spectra. For the normal branch spectra the continuum emission is well fitted with two thermal components: a soft black-body and a disk multi-colour black-body. The black-body emission represents the bulk of the emitted power in each spectrum. It is likely due to strong reprocessing in an optically thick environment formed at $R_{\rm sph}$, where disk inflow is mainly inflated by internal radiation pressure. This picture is consistent with the observed low temperature, large radius and super-Eddington luminosity \citep[see][]{shen15, soria16, Urquhart2016a,guo19}.
The hotter component, that we fitted using a \textsc{diskbb}, dominates the emission above 2 keV. Although its origin is still debated, it might come from internal hard X-ray emission, which has been inefficiently reprocessed, or simply scattered along our line of sight. Alternatively, it can be continuum emission (bremsshtrahlung and/or Comptonization) from an extended optically thin plasma where the emission lines are produced, or a tail of the black-body emission which has been Compton up-scattered in a coronal environment around the photosphere.  
In addition to the continuum, we added multiple emission and absorption lines derived by the combined averaged PN / RGS analysis to mimic the emitting and absorbing plasmas found in \citetalias[see][]{pinto21}.
Their shifts suggest different Doppler motions in several states as a consequence of a velocity field which changes depending on the launching site and on the geometry of the system. This seems supported by correlations among the parameters of the emission lines and the underneath hard X-ray flux (see Table \ref{tab:fitresults}). However, given the limited energy resolution of the EPIC we are not able to distinguish between a varying ionisation state of the plasma, the effect of a different line broadening for the different ionised species, a complex absorption/emission pattern. A thorough study of the lines is left to a dedicated forthcoming study.
For consistency the same spectral model has been applied also to the dipping branch spectra, although the physical conditions
in and out of the dips might be different.

\subsection{The normal branch: from SSUL to SUL}
\begin{figure}
\centering
\includegraphics[width=\columnwidth]{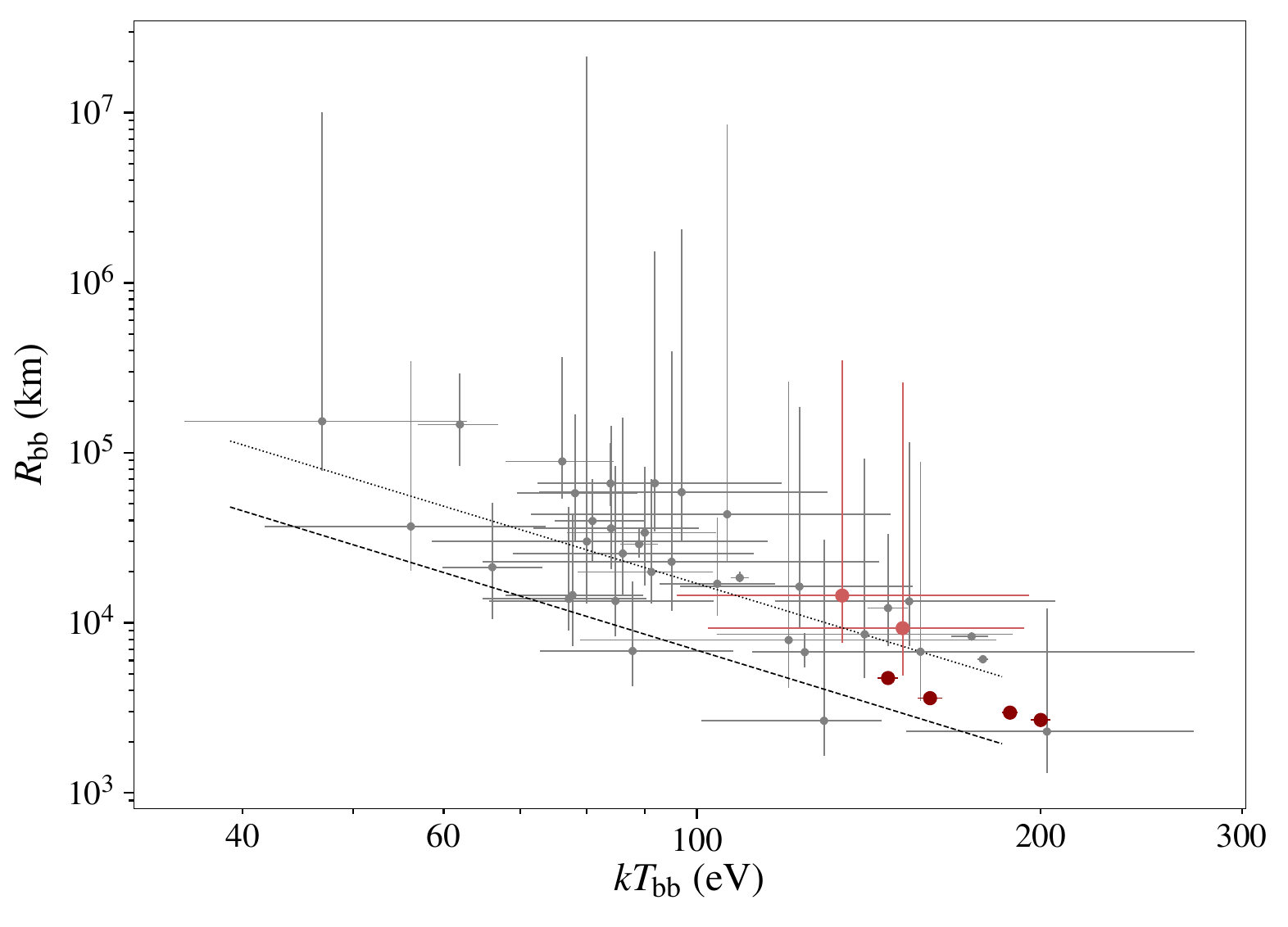}
\vspace{-0.5cm}
\caption{Radius and temperature correlation from the a sample of ULS \citep[from][]{Urquhart2016a}. The light red points are from two observations of \src\ studied in that paper. Our new measurements for the four spectra of the normal branch are dark red coloured. The dotted and the dashed lines show expected relations for a 10\,\msun BH and for a 1.4\,\msun NS according to the calculations of \citet{soria16}.}
\label{fig:plots_ktbbRbb}
\end{figure}

In the normal branch, although the luminosity of the \textsc{bbody} flux component remains almost constant, $kT_{\rm bb}$ and $R_{\rm bb}$ vary clearly in anti-correlation. From SSUL to SUL3 the $kT_{\rm bb}$ increases from 0.14 keV to 0.2 keV, while $R_{\rm bb}$ decreases from 4.7 to 2.7\,$\times$\,10$^3$ km (see Table \ref{tab:fitresults}). At variance with classical Galactic XRBs, where higher temperature and flux of a thermal component are usually explained as an increase in the local accretion rate  in SSUL ULXs that may not be the case. 
The observed emitting region dissipates locally at \rsph\ and, at the same time, reprocesses and re-emits part of the internal hard X-ray radiation. 
The fraction of the reprocessed power might strongly depend on the subtended angle, on how internal radiation is beamed, and on the disk torus height. 
According to \citet{poutanen07}, a super-Eddington disk has a photo-sphere, where the optical depth is close to unity, at a certain radius $R_{ph}$, which is proportional to the mass accretion rate $R_{ph}$ $\propto$  $\dot{m}_0^{2/3}$, where $\dot{m}_0$\,=\,$\dot{M}_0$/$\dot{M}_{Edd}$ is the dimensionless accretion rate. If the black-body radii derived from the fits are linked to this physical radius, then we estimate a decrease in the accretion rate by $\sim$\,40\% from the SSUL to the SUL3 region (see Table \ref{tab:fitresults}). 
On the other hand, the corresponding black-body temperature, is anticorrelated with $\dot{m}_0$ as $T_{\rm sph} \propto \dot{m}_0^{-1/2}$. Considering in particular the outflowing photosphere model of \citet{soria16}, our results fit well into  the vast sample of supersoft ULX sources analysed by \citet{Urquhart2016a}, when we restrict to the normal branch results. Extending the validity of this model to the dipping branch spectra could, in fact, be misleading. As it will be discussed in the next section, the mechanism behind the dips is likely not a smooth variation of the accretion rate, but a sudden change either in the observing geometry, which significantly covers part of the emitting regions, or an abrupt change in the overall accretion rate. In both cases, the spectral shape of the softer component depends on additional conditions besides the $\dot{m}_0$ value. Therefore, we inserted only the new points of \src\ obtained from our four normal branch spectra to this sample  (see Fig.\,\ref{fig:plots_ktbbRbb}). In the same plot, we reproduced the expected $kT_{\textrm {bb}}$ versus $R_{\textrm {bb}}$ relation for a 10\,\msun\ BH and for a 1.4\,\msun\ NS according to the calculations of \citet{soria16}. Our new values appear compatible with the overall trend found by \citet{Urquhart2016a} and suggest in the outflowing scenario of \citet{soria16} either a massive NS (which is likely given the high accretion rate) or a light stellar-mass BH, though this runs contrary to predictions given in \citet{yao19}, who estimated a BH mass in the range 15--60  \msun\ based on the fit result of a disk irradiation model \citep{maier12} applied to the  UV/optical spectrum of the source.

According to this scenario highest accretion rate would be thus reached in SSUL state, even though the bolometric luminosity increases going towards SUL3. The observed luminosity should scale as the logarithm of the accretion rate, so part of the luminosity in SSUL should be missed. There are many channels where this extra-luminosity might indeed hide: stronger emission in the UV regime (as the black-body temperature decreases, the fraction of the total power emitted out of band correspondingly increases, so it becomes more important to cover also the line and edge structures of the UV band to retrieve its proper measure), or transfer into kinetic power of outflows (we already noted that absorption lines seem stronger in SSUL state), or geometric effects (i.e. a change in the funnelling angle, where the region responsible for the harder emission become out of sight in SSUL state, which might also be linked to its compactness in the HID).

\citet{guo19} interpreted the SSUL/SUL transition according to the thick-disk model of \citet{gu16}, where the key physical parameter is the changing geometrical configuration of the system. This model also predicts the highest accretion rate in the SSUL regime where the inner regions are more obscured.
In a possible alternative scenario, the increasingly brighter hard tail, from SSUL to SUL, could be produced by an increase in the accretion rate which boosts the emission from the inner super-Eddington regions, followed by further scattering through the outer disk / wind (\citealt{Walton2020,gurpide21b}). This has been invoked to explain the lower pulse fraction at higher X-ray fluxes of some ULXs (\citealt{Mushtukov2021,Robba2021}).

\subsection{The origin of the dips}
Few ULXs show dips, e.g. NGC 55 ULX-1 \citep{stobbart04} and NGC 5408 ULX-1 \citep{grise13}, that also have in common with \src\ a soft spectral peak ($<$ 1 keV) and the presence of local emission features \citep{sutton15} and strong absorption lines from the wind (\citealt{Pinto2016,Pinto2017}). 

Similarly to what is observed in Galactic accreting binaries, the presence of dips might indicate that we are seeing these systems at high inclination angle. However, in classical Galactic XRBs, dips are caused by transient obscuration of the X-rays from an accretion bulge, or an outer disk structure \citep{white82}. They appear at the same orbital phases \citep[more rarely as transient phenomena, see][]{galloway16}, and they are physically connected with a spectral hardening caused by an increase in soft X-ray absorption \citep{dai14}. Presence of dips in these disk-fed, short period, systems indicate an inclination angle between 65\degmark\ and 85{\degmark}. 

In high-mass X-ray systems, dips are more rarely observed, they are not generally periodic, though they are often associated with spectral hardening \citep[see e.g.][]{naik11}. 
However, in \src\ the lack of any spectral hardening and variation in the absorption column, and the ultra-soft spectrum observed also in the dip phase, suggest that if dips are due to occultation, the covering medium should be extremely opaque (see Sect. \ref{sect:occultations}). 
An alternative scenario involving the presence of a NS and the onset of the propeller effect is discussed in Sect.\ref{sect:propeller}.

\subsubsection{The occultation scenario} \label{sect:occultations}

There is a strong hint from all the observations performed until now on this source that dips are not observed when the source is in the SSUL state, but only in the SUL state \citep[see, e.g., ][]{feng16}. If the passage from the SSUL to SUL state is driven by a long-term (of the order of days) variation in the accretion rate, then there must be a feed-back mechanism that \emph{at the same time} alters the properties of the emitting regions and the occulting clouds. In \citetalias{pinto21} and \citetalias{alston21} an increase in $\dot{m}$ was invoked to explain both the rise in the observed X-ray luminosity and the appearance of the dips, the latter being produced by higher radiation pressure pushing more optically-thick material into our the line of sight (see also \citealt{gurpide21b}). 

The ionisation parameter of the plasma in the wind is defined as $\xi=L_{\rm ion}/(n_{\rm H}R^2)$, where $L_{\rm ion}$ is the ionising luminosity, $n_{\rm H}$ the hydrogen number density and $R$ the distance from the ionising source. Assuming the numbers derived in \citetalias{pinto21}, we would obtain a wind photosphere $R=10^{12}$\,cm, from which together with the measured $0.17c$ wind velocity provides a travel time $t=R/v=200$s. This is short enough to produce the dips by obscuring the innermost regions. We note that the wind photosphere is comparable to size of the region producing the peaks in the PDS, if we assume Keplerian motion \citep{alston21}. A similar value is obtained if we calculate the maximum distance of the wind by assuming that the cloud size does not exceed its distance, i.e. $\Delta R \lesssim R$, and the column density $N_{\rm H}\approx n_{\rm H} C_{\rm V} \Delta R$ where $C_{\rm V}$ is the volume filling factor. Assuming a conservative $C_{\rm V}=0.1$ and other wind parameters from \citetalias{pinto21}, we obtain $R_{\rm max}=L_{\rm ion} C_{\rm V} /(N_{\rm H}\xi)\sim10^{12}$cm. This is consistent with the wind photosphere predicted for the Galactic super-Eddington accretor SS 433 (see, e.g., \citealt{Fabrika2004}).

Alternatively, following \citet{Lamer2003}, we can rewrite the hydrogen number density as $n_{\rm H}=N_{\rm H} / t_{\rm cross} \sqrt{(R/GM)}$, where $t_{\rm cross}$ is the crossing time and $M$ is the mass of the compact object. We find that distances of $\sim10^{12}$cm would be associated with a compact object of mass $1-10$\,\msun\ and a crossing time between 100\,s and a few hours (similar to the dips elapsed times).

We note that some dips are \emph{structured}, i.e. consisting of multiple events, as suggested by weak re-brightening episodes lasting a few hundreds of seconds. This indicates that the obscuring structure is not uniform but allows a momentary sight of the inner disk. 

\subsubsection{The propeller scenario} \label{sect:propeller}
Another possibility to explain dips is the onset of a propeller state.
We did not detect any significant coherent signal from our data, which would establish the compact object as an X-ray pulsar.
This was, however, expected, given that most of the observed radiation comes from regions distant
from the compact object and most of the hard X-ray radiation produced in the inner regions is 
strongly reprocessed (thus, possibly washing away any coherent signal, especially if short). Assuming the compact object were in any case a strongly  magnetized NS, a sufficiently low accretion rate 
would lead to an increase of the inner disk radius beyond the magnetospheric radius and, as a result, the inner flow would be stopped. 
We observe dips in SUL2/SUL3 which thus correspond to the lowest accretion rate states. This scenario is consistent  with the model by \citet{soria16}, which also explains the radius-temperature relation of the \textsc{bbody} component observed in the normal branch.  It also implies that the transition from fainter SSUL to brighter SUL states correspond to epochs in which the wind cone drops owing to a lower accretion rate thereby exposing our line of sight to the bright and hard inner disk regions.
The passage from the accretion phase to the “weak propeller” has a time-scale of around 1 ks. Here we observe that the harder component, which in our interpretation should come from illumination and reprocessing of the internal disk but also from the accretion column, is strongly hampered. Accretion flow, once threaded by the magnetic field lines, can be considered in free-fall, thus turning off very quickly the hard component flux.

The fast variability observed during dips looks remarkably like the one shown by the Hiccup Galactic pulsar, IGR\,J18245-2452 \citep[see Fig.1 in][]{ferrigno14}, although in that case dips were followed by spectral hardening.
For \src\ two key differences would be at play: super-Eddington accretion and a (likely) high inclination angle (no direct view of the magnetosphere). 

Dramatic changes in the observed flux of another pulsating ULX (PULX), M\,82 X-2, has also been explained with the propeller scenario \citep{tsygankov16}. In that case, the accretor is a 1.37 s accreting pulsar, and the source is observed to switch between two luminosity states at $\sim$1.0 $\times$ 10$^{40}$ erg s$^{-1}$ and  $\sim$2.8 $\times$ 10$^{38}$ erg s$^{-1}$. Assuming the propeller state sets in when the co-rotation radius ($R_{\rm co} = (GMP^2/4\pi^2)^{1/3}$, where $P$ is the NS spin period) is close to the magnetospheric radius ($R_{\rm mag}=k \dot{M}^{-2/7} \mu^{4/7} (2GM)^{-1/7}$, where $k$ is a geometrical  constant factor and $\mu$ is the magnetic dipole moment), and the mass accretion rate corresponds to the threshold of the minimum luminosity, an estimate on the pulsar magnetic field has been derived close to $B \simeq$ 10$^{13}$G. Similarly, for the PULX NGC 7793\,P13 a magnetic field ranging between $1-5\times$10$^{12}$ G was obtained based on the observed spin-up rate and torque model \citep{furst16,israel17}.

In the case of \src\ we have no clue on the nature of the compact object, as no flux pulsation has ever been detected. To estimate a possible period, we shall make use of the concept of the limiting luminosity ($L_{\rm lim}$) of a pulsar before entering the propeller stage \citep{stella86,campana02}. 
Assuming in the following a standard NS of 1.4 \msun\ mass ($M_{1.4}$) and 10$^6$ cm ($R_6$) radius, the limiting luminosity is defined as:
\begin{equation}
L_{\rm lim} \approx \frac{GM\dot{M}_{\rm lim}}{R} 
\approx 3.9 \times 10^{37} \, \xi^{7/2} \, B_{12}^2 \, P^{-7/3} M_{1.4}^{-2/3} R^5_6\  \ \ {\mathrm{erg~s}}^{-1}.
\end{equation}
where $\xi$ is a geometrical factor that corrects for the disk geometry the classical Alfven radius derived under spherical accretion \citep[we will assume $\xi$\,=\,0.5, see][]{campana18}; $B_{12}$ is the NS magnetic field in units of 10$^{12}$ G. At the first order, we assume that such luminosity is reached in the SSUL3 state ($L_{\rm max} \sim 2.5\times10^{39}$ \ergsec) as this is the HID region where the source more often resided before dipping.
To estimate the \emph{deep dip} flux we extracted EPIC spectra only for the time intervals where the dip reached the lowest count rate ($<$ 0.1 cps).
As expected, the spectral shape remained consistent with the D1 spectrum, while the deep dip flux is 20$\pm$10 \% lower than in D1. 
We thus obtained $L_{\rm dip} \sim 2.3\times10^{38}$ \ergsec.  In propeller the observed luminosity is decreased by a factor $R_{\rm co}/R$, which translates into a dependence on the spin period \cite[see][]{campana02}:
\begin{equation}
\frac{L_{\rm lim}}{L_{\rm dip}} \approx R_{\rm {co}}/R_{\rm NS} =  
\left( \frac{GMP^2}{4\pi^2R^3}  \right)^{1/3} 
\approx 170 \, P^{2/3}.
\end{equation}
Given our estimate $L_{\rm lim} / L_{\rm dip} \sim$\,11, we derive a spin period of $\sim$\,16 ms and $R_{\rm co}$\,=\,107 km. Because the 
$L_{\rm lim}$ value is likely a lower limit, also the derived spin period and co-rotation radius values should be considered as lower limits.

$L_{\rm dip}$ is only marginally above the Eddington limit for a 1.4\,\msun\ NS. The fraction of luminosity advected above Eddington scales as a logarithmic power of the total luminosity \citep[see Eq.6 in][]{poutanen07}, thus the residual advected fraction should be less than 10\%. The fraction of power dissipated in the disk out of the corotation radius can thus be expected to be radiatively efficient and directly linked to the corotation radius. We also assume that the remaining kinetic power is  efficiently dissipated at the disk boundary (this assumption is tacitly accepted in all works using this formula). The energy dissipated during the accretion phase is thought to be released very close to the NS. Although we do not know how a super-Eddington flow is channeled into the polar caps of the NS, we note that this scenario applied to the cases of known PULXs seem to give consistent results \citep[see][]{tsygankov16}.
To obtain also an estimate on the possible NS magnetic field value, we will now assume that $R_{\rm mag}$ $\sim$ $R_{\rm co}$.
At $R_{\rm co}$ and at the limiting luminosity, the magnetospheric radius expression is:
\begin{equation}
R_{\rm mag} \approx 750 \, \xi \, B_{12}^{4/7} \, R_6^{10/7} \, L_{39}^{-2/7} \, M_{1.4}^{1/7} \ \  {\mathrm{km}}.
\end{equation}
Taking $\xi$\,=\,0.5 and $L_{39}$\,=\,2.5, we derive a field B\,$\approx$\,2$\times$10$^{11}$ G.
Interestingly, this rough estimate is close, but below, the \rsph\ estimate derived by \citepalias{pinto21} of $\sim$ \,70 $R_g$ ($\sim$\,140 km for a $M_{1.4}$ NS).
Recent works from \citet{takahashi17} and \citet{chashkina17} provide ad-hoc prescriptions for the magnetospheric radius expression in the super-Eddington regime. We confirm that our first-order estimates are not significantly changed by adopting these more refined formulas. 

The inferred moderate magnetic field of \src\ is within the range of values derived from the whole PULX sample \citep{king20}. Galactic high-mass X-ray pulsars show B field higher by up to one order of magnitude \citep[but we also know intermediate B $\sim$\,10$^{11}$ G pulsars like GRO J1744-28,][]{dai15}, though their long-term accretion rate is significantly lower than in ULXs. It is still debated how such extreme accretion rates could change the strength and topology of NS magnetic fields \citep{choudhuri02,cumming04,lovelace05}; in some cases a significant decrease in the B-value was observed even during an outburst in a strongly magnetized pulsar \citep[see e.g.][]{cusumano16}. 
The inferred low spin period suggests that the NS could have already been significantly spun-up in the past as most PULXs show periods $\sim$ 1 s. This could be possible only if the accretion rate remained continuously super-Eddington for a sufficiently long time. The sizes and energetics of inflating super-bubbles found around many known ULXs confirm that high accretion rates have been sustained for long periods (up to 10$^{5-6}$ years, \citealt{Pakull2010,soria21}). 

\section{Conclusions}
After two decades of study, state transitions in ULXs and the physics of the super-Eddington accretion are still subjects of lively debate. We have undergone a large campaign with \xmm\ to understand the spectral transitions in the changing-look NGC 247 ULX-1, the brightest known ULX to switch between the supersoft and soft ULX states. 
Here we summarise our main results:
\begin{itemize}
    \item We obtained a complete characterisation of the spectral states in \src\ resolving the temporal spectral variations of the source into a limited number of spectral states in the HID. In this diagram, the SSUL state covers a compact and well limited region, whereas the SUL state shows a significantly higher spectral variability. The source spends a comparable time in these states. 
    \item We interpreted the softer thermal emission as produced from the outflowing photosphere at the $R_{\rm sph}$, because the radius and temperature of this component  anti-correlate; the hard component flux is the main driver of the motion of the source along the HID.

    \item The dips are observed only in the SUL state and, therefore, their presence appears linked to a threshold accretion rate value. As the source passes along the dipping branch all the emission flux components progressively diminish, but the hard component is comparatively more strongly reduced; the rapid flux decrease and line energy shift of the emission lines during dips indicate that they are produced in the inner regions.
    \item Two possible scenarios can explain dips and state transitions:
    \begin{itemize}
        \item If the accretion rate \emph{increases} from SSUL to SUL, then occultation of the emitting regions from a distant structure is the most likely scenario. A change in the accretion rate would indeed modify the \rsph\ position, and the overall geometry under which the source is observed. The time-scale, if the phenomenon is connected with the outflowing wind, is consistent with the observed ingress-egress duration; quasi-periodic features in the power spectrum indicate the occulting radius $\geq$10$^{4} R_g$.\vspace{0.2cm}
        
        \item If the accretion rate \emph{decreases} from SSUL to SUL, then we proposed the onset of a  propeller state. The lower accretion rate in the upper SUL state is mostly supported by the $R_{\rm bb}-T_{\rm bb}$ anti-correlation of the softer component, which is in agreement with  expanding photosphere model of \citet{soria16}. Assuming $R_{\rm mag} \sim R_{\rm co}$,  we estimated a relatively low spin period of 16 ms and a NS magnetic field $\approx$ 2\,$\times$\,10$^{11}$ G.
    \end{itemize}
\end{itemize}

\section*{Acknowledgements}
We  thank R. Urquhardt for having us kindly provided
the original data from \citet{Urquhart2016a} used 
for our Fig. \ref{fig:plots_ktbbRbb}. AD, MDS, FP and EA acknowledge financial contribution from the agreement ASI-INAF n.2017-14-H.0 and INAF mainstream.  EA acknowledges funding from the Italian Space Agency, contract ASI-INAF n. I/004/11/4. TPR thanks the STFC for funding as part of the consolidated grant award ST/000244/1.

\section*{Data availability}
All the data used in this article are publicly available from  ESA \xmm\ Science  Archive  (XSA; \url{https://www.cosmos.esa.int/web/xmm-newton/xsa}) and NASA HEASARC archive   (\url{https://heasarc.gsfc.nasa.gov/}). Light curves and spectra are available from the author upon request.
\bibliographystyle{mnras}
\bibliography{refs}
\bsp
\label{lastpage}
\end{document}